\documentclass[preprint,review,11pt,leqno]{elsarticle}
\usepackage[top=2.5cm,bottom=2.5cm,left=2.5cm,right=2.5cm]{geometry}
\usepackage{float}
\usepackage{graphicx}
\usepackage{grffile}
\usepackage{multirow}
\usepackage{booktabs}
\usepackage{tabularx}
\usepackage{footnote}
\usepackage{amsmath} 
\usepackage{color,amsmath}
\usepackage{color}
\usepackage{comment}
\usepackage{amssymb}
\usepackage{etoolbox}
\usepackage[hypcap=false]{caption}
\usepackage{hyperref}
\usepackage{subcaption}
\usepackage[dvipsnames]{xcolor}
\usepackage{tikz}
\usepackage{longtable}
\usepackage{setspace}
\usepackage{natbib}
\biboptions{compress}

\newcommand{\defineSpecies}[2]{\csdef{spec@#1}{#2}}
\newrobustcmd{\spec}[1]{\csuse{spec@#1}}
\defineSpecies{h216o}{H$_2^{~16}$O}
\defineSpecies{h2s}{H$_2$S}
\defineSpecies{h217o}{H$_2^{~17}$O}
\defineSpecies{h218o}{H$_2^{~18}$O}
\defineSpecies{h2o}{H$_2$O}
\defineSpecies{d216o}{D$_2^{~16}$O}
\defineSpecies{d217o}{D$_2^{~17}$O}
\defineSpecies{d218o}{D$_2^{~18}$O}
\defineSpecies{d2o}{D$_2$O}
\defineSpecies{hd16o}{HD$^{16}$O}
\defineSpecies{hd17o}{HD$^{17}$O}
\defineSpecies{hd18o}{HD$^{18}$O}
\defineSpecies{hdo}{HDO}
\defineSpecies{h2no}{H$_2~^n$O}
\defineSpecies{12c2}{$^{12}$C$_2$}
\defineSpecies{14nh3}{$^{14}$NH$_3$}
\defineSpecies{48ti16o}{$^{48}$Ti$^{16}$O}
\defineSpecies{h232s}{H$_2^{~32}$S}
\defineSpecies{so2}{SO$_2$}
\defineSpecies{90zr16o}{$^{90}$Zr$^{16}$O}
\defineSpecies{12c2h2}{$^{12}$C$_2$H$_2$}
\defineSpecies{h3+}{H$_3^+$}
\defineSpecies{h2d+}{H$_2$D$^+$}
\defineSpecies{d2h+}{D$_2$H$^+$}
\defineSpecies{16o}{$^{16}$O}
\defineSpecies{17o}{$^{17}$O}
\defineSpecies{18o}{$^{18}$O}
\defineSpecies{arno+}{Ar$\cdot$NO$^{+}$}
\defineSpecies{16o3}{$^{16}$O$_3$}
\defineSpecies{nh3}{NH$_3$}
\defineSpecies{hd}{HD}
\defineSpecies{ocs}{OCS}
\defineSpecies{13cs2}{$^{13}$CS$_2$}
\defineSpecies{co2}{CO$_2$}
\defineSpecies{ch3f}{CH$_3$F}
\defineSpecies{n2o}{N$_2$O}
\defineSpecies{co}{CO}
\defineSpecies{ch4}{CH$_4$}
\defineSpecies{n2}{N$_2$}
\defineSpecies{h2}{N$_2$}
\defineSpecies{d}{D}
\defineSpecies{h212c16o}{H$_2^{~12}$C$^{16}$O}
\defineSpecies{h3o+}{H$_3$O$^+$}
\defineSpecies{h2co}{H$_2$CO}

\newcommand{\icm}{cm$^{-1}$}

\newcommand{\be}{\begin{equation}}
\newcommand{\ee}{\end{equation}}

\newcommand{\cmimolec}{cm molecule$^{-1}$}
\newcommand{\ea}{{\it et al.}}

\newcommand{\ie}{{\it i.e.}}

\newcommand{\mc}{\multicolumn}

\renewcommand{\thetable}{\arabic{table}}

\newcommand{\PreserveBackslash}[1]{\let\temp=\\#1\let\\=\temp}
\newcolumntype{C}[1]{>{\PreserveBackslash\centering}p{#1}}
\newcolumntype{R}[1]{>{\PreserveBackslash\raggedleft}p{#1}}
\newcolumntype{L}[1]{>{\PreserveBackslash\raggedright}p{#1}}

\newcommand{\nbEl}{5029}
\newcommand{\nbTr}{19\,831}
\newcommand{\nbNonRedTr}{16\,596}
\newcommand{\nbValTr}{16\,403}
\newcommand{\nbSr}{43}

\begin{document}

\title{
An improved rovibrational linelist of formaldehyde, \texorpdfstring{\spec{h212c16o}}{H212C16O}}

\author[UCL]{Afaf R. Al-Derzi}
\author[UCL]{Jonathan Tennyson\texorpdfstring{\footnote{To whom correspondence  should be addressed; email: j.tennyson@ucl.ac.uk}}}
\author[UCL]{Sergei N. Yurchenko} 

\author[bologna]{Mattia Melosso}
\author[bologna]{Ningjing Jiang}
\author[bologna]{Cristina Puzzarini}
\author[bologna]{Luca Dore}

\author[ELTE]{Tibor Furtenbacher} 
\author[ELTE]{Roland T\'obi\'as}
\author[ELTE]{Attila G. Cs\'asz\'ar}

\address[UCL]{Department of Physics and Astronomy, University College London, 
Gower Street, London WC1E 6BT, United Kingdom}
\address[bologna]{Dipartimento di Chimica ``Giacomo Ciamician'', Universit\`a di Bologna, Via F.~Selmi 2, 40126 Bologna, Italy}
\address[ELTE]{ELKH-ELTE Complex Chemical Systems Research Group and 
ELTE E\"otv\"os Lor\'and University,
Institute of Chemistry, Laboratory of Molecular Structure and Dynamics,
H-1117 Budapest, P\'azm\'any P\'eter s\'et\'any 1/A, Hungary}

\begin{abstract}
Published high-resolution rotation-vibration transitions of \spec{h212c16o},
the principal isotopologue of methanal, are analyzed using the MARVEL 
(Measured Active Rotation-Vibration Energy Levels) procedure. 
The literature results 
are augmented by new, high-accuracy measurements of pure rotational transitions
within  the ground, $\nu_3$, $\nu_4$, and $\nu_6$ vibrational states.
Of the \nbNonRedTr\ non-redundant transitions processed, which come from 
\nbSr\ sources including the present work,
\nbValTr\ could be validated, providing \nbEl\  empirical energy
levels of \spec{h212c16o}\  with statistically well-defined uncertainties. 
All the empirical rotational-vibrational energy levels determined are used to 
improve the accuracy of ExoMol's AYTY line list for hot formaldehyde. 
The complete list of collated experimental transitions, 
the empirical energy levels determined, as well as the extended 
and improved line list  are provided as Supplementary Material.
\end{abstract}

\maketitle


\newpage
\section{Introduction}
Formaldehyde, formally methanal, HCHO, is the simplest aldehyde.
In the gas phase, formaldehyde is considered to be 
carcinogenic \cite{94Mclaughl,10NiWo}.
Since formaldehyde is a trace species in the earth's atmosphere, formed 
as a result of photo-oxidation or through incomplete biomass burning,
its carcinogenic nature motivated the development of spectroscopic
techniques for detecting it in trace quantities \cite{20HeLiFe}.

The photochemistry and photophysics of formaldehyde, involving
several excited electronic states and isomerization processes, leading to the 
products H$_2$ + CO or H + HCO, have been investigated in considerable detail 
\cite{75MoWa,83ClRa,83MoWe,00Wayne,04ToLaLeCh,09ZhMaMoBr},
both experimentally and theoretically.
The $\tilde{\rm X}\,^{1}A_1$ and $\tilde{\rm A}\,^{1}A_2$ 
electronic states of formaldehyde have been utilized
in two-photon stimulated emission pumping (SEP) experiments to derive, for example,
a complete set of 27 normal-mode vibrational constants characterizing the 
$\tilde{\rm X}\,^{1}A_1$ ground electronic state \cite{84ReFiKiDa}.
Thus, in 1984, Field and co-workers \cite{84ReFiKiDa} could legitimately claim 
that the ``unique characteristics of SEP have enabled us to describe the 
$E \leqslant 9300$ \icm\  rotation-vibration structure of H$_2$CO 
$\tilde{\rm X}\,^{1}A_1$ at
a level of detail and completeness which we believe is without precedent
for a four-atomic polyatomic molecule''.
Nevertheless, though a huge achievement,
the accuracy attainable through these spectroscopic constants was a mere 3 \icm,
hardly acceptable by present-day standards.

Over the years,
beyond anharmonic (quartic) force field representations \cite{11Csaszar}
of the $\tilde{\rm X}\,^{1}A_1$ PES of formaldehyde
\cite{93MaLeTa,95CaPiHa,96BuMcSi,18MoMaRiAg}, 
several global PESs have been developed for the [C,H,H,O] system \cite{09ZhMaMoBr,82JeBu,81GoYaSc,03JaCh,04ZhZoHaBo,08KoWaBrBo,09UlBeLeGr,11YaYuJeTh}.
Part of the interest in a global ground-electronic-state ($S_0$) PES 
stems from the existence 
of further important minima, $cis$- and $trans$-H--C--OH, 
hydroxycarbene \cite{81GoYaSc,08KoWaBrBo}, on it besides that 
corresponding to formaldehyde. 
The $trans$-hydroxycarbene minimum is some 18\,000 \icm\   \cite{08ScRePiSi}
above the  H$_2$CO minimum (the global minimum on $S_0$ belongs to CO + H$_2$ \cite{08ScRePiSi}),
separated by a transition state of 10\,400 \icm, measured from the hydroxycarbene side,
and through enhanced deep tunneling \emph{trans}-hydroxycarbene 
rearranges to formaldehyde
with a half-life of only two hours \cite{08ScRePiSi}.
The CO$\cdot$H$_2$ complex has also been the topic of 
interesting dynamical and spectroscopic studies \cite{12JaMcSz,17PaSzCs}.


Formaldehyde has been observed in different flames \cite{1817Davy,15NaKoLaBr},
most famously in cold flames, first by Sir Humphry Davy \cite{1817Davy} in 1817.
The spectrum of formaldehyde has been used for its time-resolved monitoring
in combustion engines \cite{17FjHeBaLe}, for which supporting spectroscopic 
studies have been performed \cite{20DiPeStHa}.

Formaldehyde is a trace species on Mars \cite{93KoAcKrMo}, as well, 
and it is well known in comets \cite{92BoCr,06MiReWoAb,11DeVeLiWe,11ViMuDiBo}.
Its infrared absorption  spectrum has been observed 
in protoplanetary disks \cite{14SaFoWaAl}.
Interstellar \spec{h212c16o}\  was detected 50 years ago \cite{70ZuBuPaSn}
and many of the formaldehyde lines are now resolved routinely 
in high-resolution spectroscopic studies of the interstellar medium \cite{93MaWoPl}. 
Interstellar formaldehyde masers have been widely 
observed \cite{80FoGoWiDo,07HoGoPa,13WaZhGa,14PaSo,21T2HoCl.H2CO}. 

Much of the laboratory spectroscopy of formaldehyde has focused on
electronic (rovibronic) transitions.
In fact, formaldehyde was the first polyatomic molecule where
the rotational subband structure of vibronic transitions 
has been successfully analyzed 
\cite{34DiKi}.
It is probably fair to say that the photodissociation of 
formaldehyde is one of the most thoroughly understood
polyatomic unimolecular reactions \cite{83MoWe,87ChFoMo}.
 
The importance of line-by-line information on formaldehyde has led to 
a large number of laboratory studies of its high-resolution 
rotational-vibrational spectra 
\cite{84ReFiKiDa,17FjHeBaLe,51LaSt,59TaShSh,60Oka,60OkHiSh,60ShTaSh,63Esterowi,63ShKoTa,64OkTaMo,68Takami,69NaKaKuMo,70KrGeShPo,70TuThTo,71NaMo,71TuToTh,72JoLoKi,72Nerf,72TuToTh,73ChGu,73ChFrJoOk,73JoMc,73Toth,75Nerf,75TaYaNaKu,77AlJoMc,77AlJoMc,77AlJoMcb,77ChGu,77FaKrMu,78DaWiBe,78Pine,79BrHuPi,79HaTi,80CoWi,81ChMi,81SwSa,82BrJoMcWo,85TiChKuHu,87NaDaRe,88ClVa,89ReNaDaJo,94ItNaTa,96BoDePoLi,03ThCaRiMu,88NaReDaJo,96BoHaGrSt,96LuCoFrCr,02BaCoHaPe,03BrMuLeWi,03PeKeFl,05StGaVeRu,06FlLaSaSh,06PeVaDa,06PeBrUtHa,07TcPeLa,07ZhGaDeHu,07SaBaHaRi,09MaPeJaBa,09PeJaTcLa,09CiMaCi,10JaLaTcGa,12ElCuGuHi,15RuHeHeFi,17MuLe,17TaAdNg},
involving a diverse set of experimental techniques, such as SEP \cite{84ReFiKiDa},
dispersed laser-induced fluorescence \cite{79HaTi},
conventional infrared spectroscopy, with some of the oldest results contained in 
Refs. \cite{69NaKaKuMo,71NaMo,73Toth,75TaYaNaKu,77AlJoMc},
tunable infrared difference frequency laser spectroscopy \cite{79BrHuPi},
and sub-Doppler laser Stark and Doppler-limited Fourier-transform spectroscopy \cite{82BrJoMcWo}.
Many of the results of these high-resolution experimental studies 
will be addressed and discussed in detail below.

During the present study our focus is exclusively on transitions 
within the electronic ground state and we perform a MARVEL 
(Measured Active Rotational-Vibrational Energy Levels) 
\cite{07FuCsTe,12FuCs,19ToFuTeCs} analysis of all the measured transitions 
of \spec{h212c16o} in its $\tilde{\rm X}\,^{1}A_1$ state. 
This involves collating all available laboratory spectra and then validating the
observed and assigned transitions. 
The collection of the validated lines is then inverted to give a set of 
accurate empirical energy levels with statistically significant associated 
uncertainties.
As part of this study, the published spectra
are augmented by  high-accuracy measurements of pure rotational transitions
within the ground, $\nu_3$, $\nu_4$, and $\nu_6$ vibrational states.

Accurate energy levels have a number of uses in
spectroscopy, kinetics, and thermochemistry.
One of the opportunities is to improve theoretical prediction of 
rotation-vibration spectra \cite{09TeBeBrCa}.
Recently, as part of the ExoMol project \cite{12TeYu},
Al-Refaie \ea\ \cite{15AlYuYaTe} computed an extensive rotation-vibration 
line list for hot formaldehyde, which they called AYTY. 
While this line list is comprehensive and reasonably accurate, its predicted line centers do not meet  high resolution standards.  
The empirical energy levels generated,
based partially on literature results and partially on the new measurements
performed as part of this study,
are used to make this line list
suitable for high-resolution observational studies.
This is achieved by a one-by-one replacement of the empirically  adjusted energy levels of the AYTY line list by the empirical rovibrational
energy levels resulting from our MARVEL-based investigation. 
Such high-accuracy line lists are required, for example,
by astronomers studying exoplanets using high-resolution Doppler-shift 
spectroscopy \cite{18Birkbyb}. 
Indeed, experience shows that failed
detections can in some cases be attributed to inaccurate line lists \cite{15HoKoSnBr}. 
For this reason the ExoMol project decided to refactor its line lists 
to include empirically determined energy levels wherever 
possible \cite{20TeYuAlCl}; this work forms part of this effort.

\begin{figure}[!b]
\centering
\includegraphics[width=0.9\linewidth]{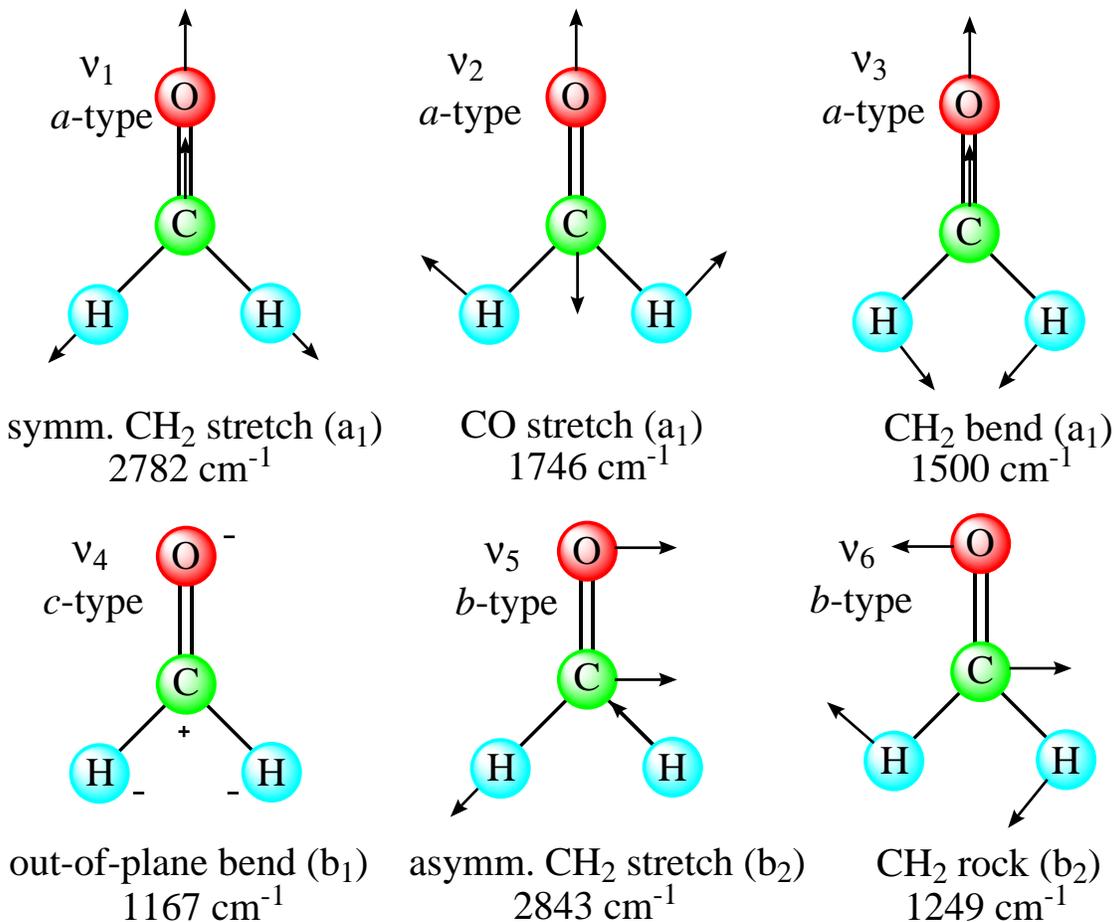}
\caption{Vibrational fundamentals of $\tilde{\rm X}\,^{1}A_1$ \spec{h212c16o},
with their symmetry species and the types of bands indicated.
The arrows representing atomic motions during the normal-mode vibrations
are not proportional to the actual displacements.}
\label{fig:H2CO_NormalModes}
\end{figure}

\section{Methods}
\subsection{MARVEL}
Details about the MARVEL 
technique \cite{07FuCsTe,12FuCs,19ToFuTeCs},
built upon the theory of spectroscopic networks 
(SNs) \cite{11CsFu,16CsFuAr,16ArFuCs}, 
and the xMARVEL code, introduced in Refs.~\citenum{19ToFuTeCs} and \citenum{20FuToTePo},
have been given in several recent 
publications \citep{19ToFuTeCs,16CsFuAr,20FuToTePo,14FuArMeCs,20ToFuSiCs}.
MARVEL has been used to treat the laboratory spectroscopic data of nine isotopologues
of water \cite{09TeBeBrCa,10TeBeBrCa,13TeBeBrCa,14TeBeBrCa,14TeBeBrCab},
as well as of
diatomics \cite{16FuSzCsBe,17McMaShSa,18McBoGoSh,19DaShJoKh,20McSyBoYu}, 
triatomics \cite{13FuSzFaCs,13FuSzMaFa,18ChNaKeBa,18ToFuCsNa}, 
tetratomics \cite{15AlFuYuTe,18ChJoFrCh,20FuCoTeYu},
and beyond \cite{11FaMaFuNe}.
Thus, it is sufficient to give only a relatively brief discussion about MARVEL here.

Of importance to the present study, the MARVEL protocol yields empirical 
rovibronic energies with well-defined provenance and 
statistically-sound uncertainties.
It always starts with the construction of a SN using the dataset of 
measured and assigned transitions collated from the literature. 
Then, a sophisticated inversion of the transitions 
is performed, yielding empirically determined rovibrational
energy levels, with associated uncertainties,
within each component of the SN.
The inversion process expedites the validation of the experimental information 
through utilization of several elements of network theory.

The xMARVEL procedure \cite{19ToFuTeCs,20FuToTePo} has been used extensively
during the present study to treat experimental rovibrational transitions of \spec{h212c16o}.
The corresponding xMARVEL input and output data files are provided as supplementary material. 

\subsection{Quantum numbers} 
The xMARVEL procedure requires as input spectral lines assigned with 
a unique set of quantum numbers. 
The equilibrium structure of formaldehyde on its ground electronic state 
has $C_{2v}$ point-group symmetry (with irreducible representations, irreps,
A$_1$, A$_2$, B$_1$, and B$_2$, which are the same for the isomorphic 
$C_{2v}$(M) molecular-symmetry group);
it is an asymmetric-top, semirigid molecule.
The vibration-rotation motions of formaldehyde can be well represented 
by a set of six vibrational normal-mode 
quantum numbers, $[v_1,v_2,v_3,v_4,v_5,v_6]$,
arranged in standard order \cite{55Mulliken}, and the three standard 
``rigid-rotor'' quantum numbers \cite{92Kroto}, $(J,K_a,K_c)$.
See Fig.~\ref{fig:H2CO_NormalModes} for a pictorial representation 
of the normal modes of \spec{h212c16o}.
Rotational states with $K_a \approx J$ which differ only in their 
$K_c$ quantum number lie close in energy and often give rise to two 
very-closely-spaced transitions, a phenomenon known as $K$-doubling or
asymmetry splitting.

The electric-dipole-allowed transitions are described as 
${\rm A}_1 \longleftrightarrow {\rm A}_2$
and ${\rm B}_1 \longleftrightarrow {\rm B}_2$.
These selection rules are derived from symmetry relations.
Namely, the two nuclear-spin isomers of \spec{h212c16o}, 
\emph{para} and \emph{ortho}, correspond to  $v_4+v_5+v_6+K_a$ being 
even (irreps A) or odd (irreps B),
respectively, while the subscripts of the irreps are 1 or 2 depending on whether
$v_5+v_6+K_a+K_c$ is even or odd, respectively.

\setstretch{1.5}
\setlength{\tabcolsep}{4pt} 
\begingroup
\captionof{table}
{Pure rotational Lamb-dip lines of \spec{h212c16o}\ measured during 
this study with 2 kHz ($6.67 \times 10^{-8}$ \icm) accuracy.$^a$}
\addtocounter{table}{-1} 
\vspace{-.3\baselineskip} 
\fontsize{10}{10}\selectfont
\begin{longtable}{rlr}\hline \hline
\endfirsthead
\hline
$\sigma_{\rm new}$/\icm & Line assignment                             & $\sigma_{\rm lit}$/\icm            \\ \hline
\endhead
\hline \mc{3}{r}{\emph{Continued on next page }} \\
\endfoot
\hline \hline
\endlastfoot
$\sigma_{\rm new}$/\icm & Line assignment                             & $\sigma_{\rm lit}$/\icm             \\ \hline
9.390\,727\,127         & ~$\nu_0, 4_{1,4} \leftarrow  3_{1,3}$ & 9.390\,727\,53(33) \cite{80CoWi}    \\   
9.694\,153\,448         & ~$\nu_0, 4_{0,4} \leftarrow  3_{0,3}$ & 9.694\,153\,31(33) \cite{80CoWi}    \\   
9.714\,646\,159         & ~$\nu_0, 4_{2,3} \leftarrow  3_{2,2}$ & 9.714\,646\,7(67) \cite{78DaWiBe}   \\ 
                        &                                             & 9.714\,646\,0(67) \cite{80CoWi}     \\   
9.719\,405\,248         & ~$\nu_0, 4_{3,2} \leftarrow  3_{3,1}$ & 9.719\,407(10) \cite{80CoWi}        \\   
9.719\,536\,101         & ~$\nu_0, 4_{3,1} \leftarrow  3_{3,0}$ & 9.719\,532\,8(33) \cite{96BoDePoLi} \\   
9.738\,339\,080         & ~$\nu_0, 4_{2,2} \leftarrow 3_{2,1}$ & 9.738\,339\,0(67) \cite{78DaWiBe}   \\   
10.034\,829\,847        & ~$\nu_0, 4_{1,3} \leftarrow  3_{1,2}$ & 10.034\,830\,00(33) \cite{80CoWi}   \\   
\end{longtable}
\normalsize 
\def\@currentlabel{\thetable}\label{tab:new_lines} 
{\setstretch{1.0}
\small
\vspace{-0.3cm}
$^a$ In the first and last columns,
the new, $\sigma_{\rm new}$, and the best previous, $\sigma_{\rm lit}$,
transition wavenumbers are reported, respectively.
The ``Line assignment'' column contains the assignments of the lines
in the form of $\nu_0, J'_{K'_a,K'_c} \leftarrow J''_{K''_a,K''_c}$, 
where     
$J_{K_a,K_c}$ denote the rotational assignment 
and $\nu_0$ refers to the ground vibrational band
both for the upper and the lower state.
\par}
\endgroup
\vspace{0.3cm}

\subsection{New rotational measurements}

In order to increase the number of available experimental data of 
\spec{h212c16o}, and thus improve our MARVEL treatment,
we have recorded a set of 89 pure rotational transitions in the ground and 
some vibrationally excited states. 
The measurements were performed using the 
frequency-modulation millimeter/submillimeter spectrometer in Bologna.
The instrumental setup of the spectrometer is detailed elsewhere 
\cite{19MeCoEsDo,20EsMeBiTa}; 
here we provide only a brief description. \\

\setlength{\tabcolsep}{4pt} 
\begingroup
\captionof{table}
{Rotational lines of \spec{h212c16o}\ measured during this 
study with 20 kHz ($6.67 \times 10^{-7}$ \icm) accuracy.$^a$}
\addtocounter{table}{-1} 
\vspace{-.3\baselineskip} 
\fontsize{10}{10}\selectfont
\begin{longtable}{r@{\hskip 0.3cm}r@{\hskip 0.1cm}R{0.15cm}L{1.3cm}
                  r@{\hskip 0.3cm}r@{\hskip 0.1cm}R{0.15cm}l
                  r@{\hskip 0.3cm}r@{\hskip 0.1cm}R{0.15cm}l}
\hline \hline
\endfirsthead
\mc{12}{c}{\tablename\ \thetable\ -- \emph{Continued from previous page}}\\ \hline 
$\sigma_{\rm new}$/\icm 
                 & \mc{3}{c}{\hspace{-0.6cm}Line assignment}              & $\sigma_{\rm new}$/\icm 
                                                                                             & \mc{3}{c}{\hspace{-0.3cm}Line assignment}              & $\sigma_{\rm new}$/\icm
                                                                                                                                                                         & \mc{3}{c}{\hspace{-0.3cm}Line assignment}              \\ \hline
\endhead
\hline \mc{12}{r}{\emph{ Continued on next page }} \\
\endfoot
\hline \hline
\endlastfoot
$\sigma_{\rm new}$/\icm 
                 & \mc{3}{c}{\hspace{-0.6cm}Line assignment}              & $\sigma_{\rm new}$/\icm 
                                                                                             & \mc{3}{c}{\hspace{-0.3cm}Line assignment}              & $\sigma_{\rm new}$/\icm
                                                                                                                                                                         & \mc{3}{c}{\hspace{-0.3cm}Line assignment}              \\ \hline
9.396\,103\,294  & ~$\nu_3, 4_{1,4}$   & $\leftarrow$ & $3_{1,3}$   & 12.098\,165\,470 & ~$\nu_4, 5_{3,3}$   & $\leftarrow$ & $4_{3,2}$   & 9.694\,153\,398  & ~$\nu_0, 4_{0,4}$   & $\leftarrow$ & $3_{0,3}$   \\ 
9.720\,532\,400  & ~$\nu_3, 4_{0,4}$   & $\leftarrow$ & $3_{0,3}$   & 12.098\,598\,600 & ~$\nu_4, 5_{3,2}$   & $\leftarrow$ & $4_{3,1}$   & 9.714\,646\,044  & ~$\nu_0, 4_{2,3}$   & $\leftarrow$ & $3_{2,2}$   \\ 
9.745\,845\,447  & ~$\nu_3, 4_{2,3}$   & $\leftarrow$ & $3_{2,2}$   & 12.129\,667\,850 & ~$\nu_4, 5_{2,3}$   & $\leftarrow$ & $4_{2,2}$   & 9.719\,405\,501  & ~$\nu_0, 4_{3,2}$   & $\leftarrow$ & $3_{3,1}$   \\ 
9.753\,700\,377  & ~$\nu_3, 4_{3,2}$   & $\leftarrow$ & $3_{3,1}$   & 12.432\,459\,250 & ~$\nu_4, 5_{1,4}$   & $\leftarrow$ & $4_{1,3}$   & 9.719\,535\,726  & ~$\nu_0, 4_{3,1}$   & $\leftarrow$ & $3_{3,0}$   \\ 
9.753\,859\,167  & ~$\nu_3, 4_{3,1}$   & $\leftarrow$ & $3_{3,0}$   & 9.366\,709\,450  & ~$\nu_6, 4_{1,4}$   & $\leftarrow$ & $3_{1,3}$   & 9.738\,339\,087  & ~$\nu_0, 4_{2,2}$   & $\leftarrow$ & $3_{2,1}$   \\ 
9.773\,007\,973  & ~$\nu_3, 4_{2,2}$   & $\leftarrow$ & $3_{2,1}$   & 9.682\,101\,847  & ~$\nu_6, 4_{0,4}$   & $\leftarrow$ & $3_{0,3}$   & 9.834\,452\,846  & ~$\nu_0, 22_{2,20}$ & $\leftarrow$ & $23_{0,23}$ \\ 
10.088\,575\,460 & ~$\nu_3, 4_{1,3}$   & $\leftarrow$ & $3_{1,2}$   & 9.697\,356\,479  & ~$\nu_6, 4_{2,3}$   & $\leftarrow$ & $3_{2,2}$   & 10.034\,829\,710 & ~$\nu_0, 4_{1,3}$   & $\leftarrow$ & $3_{1,2}$   \\ 
11.739\,822\,940 & ~$\nu_3, 5_{1,5}$   & $\leftarrow$ & $4_{1,4}$   & 9.696\,151\,729  & ~$\nu_6, 4_{3,2}$   & $\leftarrow$ & $3_{3,1}$   & 10.248\,367\,590 & ~$\nu_0, 9_{2,7}$   & $\leftarrow$ & $10_{0,10}$ \\ 
12.129\,667\,850 & ~$\nu_3, 5_{0,5}$   & $\leftarrow$ & $4_{0,4}$   & 9.696\,255\,862  & ~$\nu_6, 4_{3,1}$   & $\leftarrow$ & $3_{3,0}$   & 10.364\,365\,470 & ~$\nu_0, 18_{3,15}$ & $\leftarrow$ & $19_{1,18}$ \\ 
12.178\,445\,400 & ~$\nu_3, 5_{2,4}$   & $\leftarrow$ & $4_{2,3}$   & 9.718\,870\,594  & ~$\nu_6, 4_{2,2}$   & $\leftarrow$ & $3_{2,1}$   & 10.448\,342\,040 & ~$\nu_0, 37_{4,33}$ & $\leftarrow$ & $37_{4,34}$ \\ 
12.192\,149\,190 & ~$\nu_3, 5_{4,2}$   & $\leftarrow$ & $4_{4,1}$   & 10.029\,447\,240 & ~$\nu_6, 4_{1,3}$   & $\leftarrow$ & $3_{1,2}$   & 10.555\,863\,330 & ~$\nu_0, 11_{1,10}$ & $\leftarrow$ & $11_{1,11}$ \\ 
12.192\,149\,190 & ~$\nu_3, 5_{4,1}$   & $\leftarrow$ & $4_{4,0}$   & 11.704\,713\,210 & ~$\nu_6, 5_{1,5}$   & $\leftarrow$ & $4_{1,4}$   & 10.896\,074\,830 & ~$\nu_0, 19_{2,17}$ & $\leftarrow$ & $19_{2,18}$ \\ 
12.194\,013\,170 & ~$\nu_3, 5_{3,3}$   & $\leftarrow$ & $4_{3,2}$   & 12.086\,508\,690 & ~$\nu_6, 5_{0,5}$   & $\leftarrow$ & $4_{0,4}$   & 11.212\,427\,070 & ~$\nu_0, 28_{3,25}$ & $\leftarrow$ & $28_{3,26}$ \\ 
12.194\,571\,510 & ~$\nu_3, 5_{3,2}$   & $\leftarrow$ & $4_{3,1}$   & 12.120\,488\,600 & ~$\nu_6, 5_{2,4}$   & $\leftarrow$ & $4_{2,3}$   & 11.733\,738\,380 & ~$\nu_0, 5_{1,5}$   & $\leftarrow$ & $4_{1,4}$   \\ 
12.232\,626\,840 & ~$\nu_3, 5_{2,3}$   & $\leftarrow$ & $4_{2,2}$   & 12.109\,817\,660 & ~$\nu_6, 5_{4,2}$   & $\leftarrow$ & $4_{4,1}$   & 12.073\,937\,490 & ~$\nu_0, 8_{2,6}$   & $\leftarrow$ & $9_{0,9}$   \\ 
12.604\,848\,190 & ~$\nu_3, 5_{1,4}$   & $\leftarrow$ & $4_{1,3}$   & 12.109\,817\,660 & ~$\nu_6, 5_{4,1}$   & $\leftarrow$ & $4_{4,0}$   & 12.099\,570\,770 & ~$\nu_0, 5_{0,5}$   & $\leftarrow$ & $4_{0,4}$   \\ 
9.384\,944\,381  & ~$\nu_4, 4_{1,4}$   & $\leftarrow$ & $3_{1,3}$   & 12.121\,054\,530 & ~$\nu_6, 5_{3,3}$   & $\leftarrow$ & $4_{3,2}$   & 12.139\,927\,050 & ~$\nu_0, 5_{2,4}$   & $\leftarrow$ & $4_{2,3}$   \\ 
9.649\,643\,518  & ~$\nu_4, 4_{0,4}$   & $\leftarrow$ & $3_{0,3}$   & 12.121\,422\,250 & ~$\nu_6, 5_{3,2}$   & $\leftarrow$ & $4_{3,1}$   & 12.145\,176\,040 & ~$\nu_0, 5_{4,2}$   & $\leftarrow$ & $4_{4,1}$   \\ 
9.670\,558\,195  & ~$\nu_4, 4_{2,3}$   & $\leftarrow$ & $3_{2,2}$   & 12.163\,422\,430 & ~$\nu_6, 5_{2,3}$   & $\leftarrow$ & $4_{2,2}$   & 12.145\,176\,040 & ~$\nu_0, 5_{4,1}$   & $\leftarrow$ & $4_{4,0}$   \\ 
9.677\,770\,970  & ~$\nu_4, 4_{3,2}$   & $\leftarrow$ & $3_{3,1}$   & 12.532\,770\,930 & ~$\nu_6, 5_{1,4}$   & $\leftarrow$ & $4_{1,3}$   & 12.150\,912\,170 & ~$\nu_0, 5_{3,3}$   & $\leftarrow$ & $4_{3,2}$   \\ 
9.677\,893\,765  & ~$\nu_4, 4_{3,1}$   & $\leftarrow$ & $3_{3,0}$   & 8.237\,600\,872  & ~$\nu_0, 21_{2,19}$ & $\leftarrow$ & $22_{0,22}$ & 12.151\,369\,510 & ~$\nu_0, 5_{3,2}$   & $\leftarrow$ & $4_{3,1}$   \\ 
9.693\,290\,175  & ~$\nu_4, 4_{2,2}$   & $\leftarrow$ & $3_{2,1}$   & 8.600\,996\,014  & ~$\nu_0, 35_{3,32}$ & $\leftarrow$ & $36_{1,35}$ & 12.187\,211\,040 & ~$\nu_0, 5_{2,3}$   & $\leftarrow$ & $4_{2,2}$   \\ 
9.949\,961\,352  & ~$\nu_4, 4_{1,3}$   & $\leftarrow$ & $3_{1,2}$   & 8.615\,829\,497  & ~$\nu_0, 10_{2,8}$  & $\leftarrow$ & $11_{0,11}$ & 12.441\,481\,610 & ~$\nu_0, 12_{1,11}$ & $\leftarrow$ & $12_{1,12}$ \\ 
11.726\,529\,220 & ~$\nu_4, 5_{1,5}$   & $\leftarrow$ & $4_{1,4}$   & 8.798\,468\,478  & ~$\nu_0, 36_{4,32}$ & $\leftarrow$ & $36_{4,33}$ & 12.538\,447\,510 & ~$\nu_0, 5_{1,4}$   & $\leftarrow$ & $4_{1,3}$   \\ 
12.045\,134\,760 & ~$\nu_4, 5_{0,5}$   & $\leftarrow$ & $4_{0,4}$   & 8.815\,102\,825  & ~$\nu_0, 10_{1,9}$  & $\leftarrow$ & $10_{1,10}$ & 12.795\,269\,680 & ~$\nu_0, 20_{2,18}$ & $\leftarrow$ & $20_{2,19}$ \\ 
12.084\,311\,370 & ~$\nu_4, 5_{2,4}$   & $\leftarrow$ & $4_{2,3}$   & 9.160\,256\,575  & ~$\nu_0, 18_{2,16}$ & $\leftarrow$ & $18_{2,17}$ &                  &                    &              &                    \\ 
12.093\,976\,360 & ~$\nu_4, 5_{4,2}$   & $\leftarrow$ & $4_{4,1}$   & 9.390\,727\,004  & ~$\nu_0, 4_{1,4}$   & $\leftarrow$ & $3_{1,3}$   &                  &                    &              &                    \\ 
12.093\,936\,030 & ~$\nu_4, 5_{4,1}$   & $\leftarrow$ & $4_{4,0}$   & 9.461\,197\,252  & ~$\nu_0, 27_{3,24}$ & $\leftarrow$ & $27_{3,25}$ &                  &                    &              &                    \\ 
\end{longtable}
\normalsize
\def\@currentlabel{\thetable}\label{tab:new_lines2} 
{\setstretch{1.0}
\small
\vspace{-0.3cm}
$^a$ $\sigma_{\rm new}$ denotes newly measured transition wavenumbers.
     The line assignments are given in the form of
     $\nu_i, J'_{K'_a,K'_c} \leftarrow J''_{K''_a,K''_c}$, where
     $\nu_i$ and $J_{K_a,K_c}$ represent the vibrational and the rotational
     assignment of the measured lines, respectively, $\nu_0$ refers to
     the ground vibrational state, while $\nu_3,~\nu_4, {\rm and ~} \nu_6$
     designate  
     vibrational fundamentals (see Fig.~\ref{fig:H2CO_NormalModes}).
\par} 
\endgroup

\vspace{0.5cm}
The radiation source is a Gunn diode (80--115~GHz) that is frequency- and 
phase-stabilized \emph{via} a phase-lock-loop and is driven by a centimeter-wave 
synthesizer referenced to a 5~MHz rubidium atomic clock. 
Spectral coverage between 240 and 420~GHz is obtained 
by coupling the Gunn diode to passive multipliers.
Then, the electromagnetic radiation is frequency-modulated at 
$f=48$~kHz and fed into a glass absorption cell (3~m optical path length) filled 
with formaldehyde vapors at pressure between 0.5 and 20~$\mu$bar; specifically, 
pressures between 0.5 and 20~$\mu$bar were used to record ground state lines with 
intensities spanning almost six orders of magnitude, while 10--15~$\mu$bar 
were used to observe the vibrational excited states.
The sample of H$_2$CO was freshly obtained before each measurement from the vapors of solid,
room-temperature paraformaldehyde.
The output radiation was finally detected by a Schottky barrier 
diode, demodulated by a lock-in amplifier set at twice the 
modulation-frequency ($2f$ scheme), filtered, and analog-to-digitally converted.

A small sub-set of measurements have been performed exploiting the Lamb-dip 
technique \cite{64Lamb}. 
In this respect, the optics of the spectrometer 
were adequately set up in a double-pass configuration (as detailed in 
\cite{20MeDoGaPu}), $f$ was set to 1~kHz, and a low-pressure of H$_2$CO was 
used (0.5~$\mu$bar).

The new measurements consist of 
(a) two complete $a$-type $^qR$-branch transitions 
    ($J = 4 \leftarrow 3$ and $5 \leftarrow 4$) within the $\nu_0$, 
	$\nu_3$, $\nu_4$, and $\nu_6$ bands, where $\nu_0$ denotes
	the ground vibrational state, 
    and
(b) about 20 ground-state $P$ and $Q$ transitions with 
	$\Delta K_a = 0$ or $2$, with integrated intensity between $10^{-23}$ and $10^{-27}$~\cmimolec.

The uncertainty of our measurements is estimated to be in the range of 2--20~kHz.
We label the seven Lamb-dip transitions of Table~\ref{tab:new_lines}
with the reference tag 21AlTeYuMe, and the others as
21AlTeYuMe\_S2, see Table~\ref{tab:new_lines2}. 

\subsection{Artificial transitions}
In the absence of transitions linking the \emph{ortho} and \emph{para} 
nuclear-spin isomers, the measured transitions form two distinct 
principal components (PC) \cite{16CsFuAr} in the SN of \spec{h212c16o}. 
As usual \cite{09TeBeBrCa} during MARVEL analyses of experimental transitions 
of species having more than one nuclear-spin isomer,
the \emph{ortho} and \emph{para} principal components are linked
using artificial transitions, colloquially called
magic numbers, taken possibly from effective Hamiltonian fits (EH) \cite{20ToFuSiCs}.

For \spec{h212c16o}, the wavenumber of the 
unmeasurable rotational transition $1_{1,1} \leftarrow 0_{0,0}$, reported in
\cite{97CaHaDe} as 10.539\,039\,1 \icm, is utilized as a magic number.
Since the very accurate empirical (MARVEL) energies of the pure rotational states
$1_{0,1}$, $2_{0,2}$, and $3_{0,3}$, that is, 2.429\,612\,60(33), 
7.286\,404\,28(47), and 14.565\,513\,07(58) \icm, respectively, are reproduced 
by the EH-predicted values, see the 
fourth column of Table 3 of Ref.~\cite{97CaHaDe},
with unsigned deviations of $8.26 \times 10^{-7}$, 
$2.20 \times 10^{-6}$, and $4.37 \times 10^{-6}$ \icm, respectively,
the average of these unsigned deviations, $2.3 \times 10^{-6}$ \icm, 
is adopted as a conservative estimate for the uncertainty of the
magic number.
In addition, 11 artificial transitions were used
to link the largest floating components to the PCs. 
These transitions form bridges between the principal components
and the floating components of disconnected higher-$J$ transitions. 
In fact, these transitions link the ground state 
to series of states with $K_a=8$, $K_a=10$, and $K_a=11$. 
The values of these artificial transitions are taken from the effective
Hamiltonian study of 88NaReDaJo \cite{88NaReDaJo}.

First-principles computations are capable of estimating
small energy splittings very accurately \cite{20ToFuSiCs}; thus, we decided to add 
the $J_{J,0/1}$ separations as virtual lines to the dataset.
They are part of a source tagged as `21virt' (see Table~\ref{tab:marveldata}).
These virtual transitions are distributed into four segments,
based on the magnitude of the splittings and thus on their assumed uncertainties.
Through these virtual lines 199 further rovibrational energy levels could be
determined:
the experimentally unavailable states of certain $J_{J,0/1}$ pairs.
\vspace{0.75cm}

\setstretch{1.3}
\setlength{\tabcolsep}{4pt} 
\begingroup
\captionof{table}[Data source segments and their characteristics for 
                the \spec{h212c16o}\ molecule]
                {Data source segments and their characteristics for 
                the \spec{h212c16o}\ molecule$^{a}$} 
\addtocounter{table}{-1} 
\vspace{-.5\baselineskip} 
\scriptsize
\begin{longtable}{lr@{--}p{1.0cm}>{\centering\arraybackslash}p{3.0cm}rrrl}
\hline \hline
\endfirsthead
\mc{6}{c}{\tablename\ \thetable\ -- \emph{Continued from previous page}}\\ \hline
     Segment tag                                   &  \mc{2}{c}{Range}         &        $A/N/V$               &        ESU  &        MSU  &        LSU  &  Recalib. Factor  \\ \hline
\endhead
\hline \mc{8}{r}{\emph{ Continued on next page }} \\
\endfoot
\hline \hline
\endlastfoot
     Segment tag                                   &  \mc{2}{c}{Range}         &        $A/N/V$               &        ESU  &        MSU  &        LSU  &  Recalib. Factor  \\ \hline
     72TuToTh \cite{72TuToTh}                      &  0.483\,28  &  0.483\,28  &         1/1/1                &   2.67e-09  &   2.67e-09  &   2.67e-09  &                   \\
     71TuToTh \cite{71TuToTh}                      &  0.141\,30  &  0.165\,74  &         3/3/3                &   3.67e-09  &   3.67e-09  &   5.34e-09  &                   \\
     81ChMi \cite{81ChMi}                          &  0.002\,373 &  0.064\,793 &        14/9/9                &   6.67e-09  &   1.00e-08  &   3.34e-08  &                   \\
     73ChGu \cite{73ChGu}                          &  0.000\,003 &  0.966\,50  &        23/21/21              &   1.00e-08  &   1.00e-08  &   3.66e-07  &                   \\
     73ChGu\_S2 \cite{73ChGu}                      &  2.415\,3   &  2.415\,3   &         1/1/1                &   3.34e-05  &   3.34e-05  &   3.34e-05  &                   \\
     68Takami \cite{68Takami}                      &  0.000\,022 &  0.001\,829 &        12/10/10              &   1.67e-08  &   3.15e-08  &   6.44e-07  &                   \\
     77ChGu \cite{77ChGu}                          &  0.000\,005 &  0.002\,305 &        39/39/35              &   3.34e-08  &   3.34e-08  &   1.33e-07  &                   \\
     96BoDePoLi \cite{96BoDePoLi}                  &  0.000\,003 &  0.708\,93  &        68/35/35              &   3.34e-08  &   1.67e-07  &   1.67e-07  &                   \\
     96BoDePoLi\_S2 \cite{96BoDePoLi}              &  0.000\,059 &  46.275     &        93/45/44              &   1.00e-06  &   1.00e-06  &   4.37e-06  &                   \\
     96BoDePoLi\_S3 \cite{96BoDePoLi}              &  0.165\,27  &  63.241     &        91/72/72              &   3.34e-06  &   3.34e-06  &   1.67e-05  &                   \\
     96BoDePoLi\_S4 \cite{96BoDePoLi}              &  34.009     &  84.711     &        21/21/21              &   1.67e-04  &   1.67e-04  &   1.67e-04  &                   \\
     21AlTeYuMe                                    &  9.390\,7   &  10.035     &         7/7/7                &   7.00e-08  &   7.00e-08  &   1.33e-07  &                   \\
     21AlTeYuMe\_S2                                &  8.237\,6   &  12.988     &        82/82/82              &   7.00e-07  &   7.00e-07  &   4.74e-06  &                   \\
     66TaEvSh \cite{66TaEvSh}                      &  0.000\,153 &  0.000\,610 &         2/2/2                &   1.00e-07  &   1.00e-07  &   1.33e-07  &                   \\
     77FaKrMu \cite{77FaKrMu}                      &  0.002\,373 &  0.082\,838 &         4/4/4                &   1.67e-07  &   1.67e-07  &   1.67e-07  &                   \\
     63Esterowi \cite{63Esterowi}                  &  0.010\,044 &  0.035\,553 &         4/4/4                &   2.50e-07  &   4.17e-06  &   1.11e-05  &                   \\
     80CoWi \cite{80CoWi}                          &  0.000\,003 &  15.039     &       127/95/94              &   3.34e-07  &   3.34e-07  &   3.15e-06  &                   \\
     17MuLe \cite{17MuLe}                          &  45.744     &  49.915     &        45/45/45              &   3.34e-07  &   3.34e-07  &   3.34e-07  &                   \\
     88NaReDaJo \cite{88NaReDaJo}                  &  0.000\,020 &  0.561\,82  &        10/10/10              &   6.67e-07  &   6.67e-07  &   1.12e-05  &                   \\
     88NaReDaJo\_S2 \cite{88NaReDaJo}              &  0.000\,005 &  10.029     &        64/64/63              &   1.00e-06  &   1.00e-06  &   1.72e-05  &                   \\
     88NaReDaJo\_S3 \cite{88NaReDaJo}              &  608.85     &  630.67     &         2/2/2                &   5.00e-05  &   1.00e-05  &   1.00e-05  &                   \\
     88NaReDaJo\_S4 \cite{88NaReDaJo}              &  922.63     &  1\,578.4   &    314\,9/312\,8/312\,8      &   2.00e-04  &   1.19e-04  &   2.01e-02  &                   \\
     88NaReDaJo\_S5 \cite{88NaReDaJo}              &  681.57     &  1\,201.9   &         9/9/9                &   5.00e-03  &   1.00e-03  &   1.00e-03  &                   \\
     03ThCaRiMu \cite{03ThCaRiMu}                  &  0.163\,02  &  0.164\,96  &         3/3/2                &   8.34e-07  &   8.34e-07  &   1.00e-06  &                   \\
     03BrMuLeWi \cite{03BrMuLeWi}                  &  27.793     &  66.778     &       136/136/136            &   1.00e-06  &   1.00e-06  &   2.74e-05  &                   \\
     09MaPeJaBa \cite{09MaPeJaBa}                  &  5.016\,8   &  30.115     &       172/172/171            &   1.00e-06  &   1.00e-06  &   1.33e-05  &                   \\
     12ElCuGuHi \cite{12ElCuGuHi}                  &  23.353     &  58.596     &        87/87/87              &   1.00e-06  &   1.00e-06  &   5.55e-06  &                   \\
     21virt                                        &  0.000\,000 &  0.000\,000 &       446/446/446            &   1.00e-06  &   1.00e-06  &   1.00e-06  &                   \\
     21virt\_S2                                    &  0.000\,001 &  0.000\,009 &       178/177/177            &   1.00e-05  &   1.00e-05  &   2.21e-04  &                   \\
     21virt\_S3                                    &  0.000\,010 &  0.000\,098 &       157/151/151            &   1.00e-04  &   1.00e-04  &   2.78e-03  &                   \\
     21virt\_S4                                    &  0.000\,100 &  0.000\,999 &       168/166/166            &   1.00e-03  &   1.00e-03  &   3.89e-03  &                   \\
     72Nerf \cite{72Nerf}                          &  1.610\,6   &  10.035     &        15/15/15              &   1.47e-06  &   1.60e-06  &   7.86e-06  &                   \\
     51LaSt \cite{51LaSt}                          &  0.245\,59  &  2.429\,6   &        18/18/18              &   1.67e-06  &   2.40e-06  &   2.03e-05  &                   \\
     56Erlandss \cite{56Erlandss}                  &  2.429\,6   &  7.528\,5   &         9/8/8                &   1.67e-06  &   1.60e-05  &   2.79e-05  &                   \\
     64OkTaMo \cite{64OkTaMo}                      &  0.000\,153 &  2.437\,1   &        70/49/44              &   1.67e-06  &   1.80e-06  &   2.66e-05  &                   \\
     97CaHaDe \cite{97CaHaDe}                      &  10.539     &  10.539     &         1/1/1                &   2.30e-06  &   2.30e-06  &   2.30e-06  &                   \\
     72JoLoKi \cite{72JoLoKi}                      &  0.000\,153 &  4.504\,1   &        37/10/10              &   3.34e-06  &   8.34e-07  &   5.00e-06  &                   \\
     78DaWiBe \cite{78DaWiBe}                      &  0.353\,87  &  12.187     &        86/82/82              &   6.67e-06  &   6.67e-06  &   2.31e-05  &                   \\
     73ChFrJoOk \cite{73ChFrJoOk}                  &  0.553\,19  &  1.979\,8   &         7/7/7                &   1.67e-05  &   1.67e-05  &   2.67e-05  &                   \\
     85TiChKuHu \cite{85TiChKuHu}                  &  4\,211.6   &  5\,224.0   &    100\,5/100\,5/100\,3      &   1.00e-04  &   1.10e-04  &   1.16e-02  &                   \\
     82BrJoMcWo \cite{82BrJoMcWo}                  &  1\,693.1   &  1\,793.4   &       248/248/248            &   4.00e-04  &   1.59e-04  &   8.99e-04  &                   \\
     79BrHuPi \cite{79BrHuPi}                      &  2\,700.0   &  3\,000.2   &    319\,3/318\,9/318\,3      &   5.00e-04  &   3.27e-04  &   4.52e-02  &                   \\
     81SwSa \cite{81SwSa}                          &  1\,707.1   &  1\,746.9   &        76/76/76              &   5.00e-04  &   6.50e-04  &   3.38e-03  &                   \\
     88ClVa \cite{88ClVa}                          &  2\,867.2   &  2\,880.4   &        24/22/22              &   5.00e-04  &   3.48e-04  &   7.23e-03  &                   \\
     94ItNaTa \cite{94ItNaTa}                      &  2\,276.2   &  2\,553.6   &       320/320/317            &   5.00e-04  &   3.91e-04  &   5.56e-03  &                   \\
     10JaLaTcGa \cite{10JaLaTcGa}                  &  1\,675.0   &  3\,089.8   &       782/781/781            &   5.00e-04  &   1.62e-04  &   6.45e-03  &  1.000\,001\,293  \\
     07TcPeLa\_S2 \cite{07TcPeLa}                  &  0.000\,006 &  10.029     &        84/84/84              &   2.10e-04  &   1.00e-04  &   2.25e-03  &                   \\
     07TcPeLa \cite{07TcPeLa}                      &  927.68     &  1\,821.6   &    402\,4/110\,1/110\,1      &   8.00e-04  &   1.04e-04  &   5.00e-03  &                   \\
     06PeBrUtHa \cite{06PeBrUtHa}                  &  2\,756.6   &  2\,864.6   &       147/143/143            &   1.00e-03  &   1.00e-03  &   3.82e-02  &                   \\
     09CiMaCi \cite{09CiMaCi}                      &  4\,350.9   &  4\,360.6   &        49/46/36              &   1.00e-03  &   1.00e-03  &   7.97e-03  &                   \\
     07SaBaHaRi \cite{07SaBaHaRi}                  &  5\,597.8   &  5\,698.3   &       424/422/418            &   1.50e-03  &   1.00e-03  &   9.08e-03  &                   \\
     17TaAdNg \cite{17TaAdNg}                      &  3\,398.7   &  3\,529.7   &       786/786/785            &   1.60e-03  &   1.11e-03  &   3.66e-02  &                   \\
     73Toth \cite{73Toth}                          &  3\,428.2   &  3\,507.4   &       299/299/298            &   3.00e-03  &   2.83e-03  &   4.23e-02  &                   \\
     77AlJoMc \cite{77AlJoMc}                      &  945.23     &  1\,540.0   &       991/922/904            &   5.00e-03  &   3.94e-03  &   4.57e-02  &                   \\
     77AlJoMcb \cite{77AlJoMcb}                    &  1\,483.0   &  1\,518.1   &        70/70/70              &   5.00e-03  &   4.14e-03  &   3.57e-02  &                   \\
     76NaYaKu \cite{76NaYaKu}                      &  2\,643.9   &  3\,011.8   &    184\,8/184\,5/171\,1      &   3.00e-02  &   1.68e-02  &   9.89e-02  &                   
\end{longtable}
\normalsize 
\def\@currentlabel{\thetable}\label{tab:marveldata} 
{\setstretch{1.0}
\small
$^{a}$ Tags denote segments used in this study. 
The column `Range' indicates the range (in \icm) corresponding to validated
wavenumbers within the transition list.
$A$ is the number of assigned transitions, $N$ is the number of 
non-redundant lines (with distinct wavenumbers or labels), and $V$ is 
the number of validated transitions obtained at the end of the xMARVEL
analysis.
In the heading of this table, ESU, MSU, and LSU denote the estimated, 
the median, and the largest segment uncertainties, respectively, in \icm.
Rows are arranged in the order of the ESUs with the restriction 
that the segments of the same data source should be listed consecutively.\\
\par}
\endgroup

\begin{table}[!b]
\caption{High-resolution spectroscopic studies on \spec{h212c16o}\   which were 
considered but not utilized during the present MARVEL analysis, 
with reason for the exclusion.}
\label{tab:marvelnodata}
\resizebox{\columnwidth}{!}{
\begin{tabular}{lcl}
\hline \hline 
Tag                          & Range / \icm  & Reason for exclusion                                                \\ \hline  
17FjHeBaLe \cite{17FjHeBaLe} & 6230 -- 6240  & 700 K emission spectrum, line parameters not provided               \\
15RuHeHeFi \cite{15RuHeHeFi} & 6547 -- 7051  & No line assignments                                                 \\
09PeJaTcLa \cite{09PeJaTcLa} & 1600 -- 3200  & Calculated line positions only                                      \\
07ZhGaDeHu \cite{07ZhGaDeHu} & 6351 -- 6362  & No line assignments                                                 \\
06FlLaSaSh \cite{06FlLaSaSh} & 3096 -- 5263  & Data not made available by the authors                              \\
06PeVaDa \cite{06PeVaDa}     & 2600 -- 3100  & No line parameters provided, data not made available by the authors \\
05StGaVeRu \cite{05StGaVeRu} & 6547 -- 6804  & No line assignments                                                 \\
03PeKeFl \cite{03PeKeFl}     & 1000 -- 2000  & Data not made available by the authors                              \\
02BaCoHaPe \cite{02BaCoHaPe} & 5600 -- 5700  & Data analysed by 07SaBaHaRi \cite{07SaBaHaRi}                  \\
96BoHaGrSt \cite{96BoHaGrSt} &  --           & Dispersed fluorescence, vibrational state data only                 \\
96LuCoFrCr \cite{96LuCoFrCr} & 7800 -- 15200 & No line parameters provided                                         \\
89ReNaDaJo \cite{89ReNaDaJo} & 890 -- 1590   & No line parameters provided                                         \\
87NaDaRe \cite{87NaDaRe}     & 1148 -- 1193  & Lines included in 88NaReDaJo \cite{88NaReDaJo}                      \\
78Pine \cite{78Pine}         & 2700 -- 3000  & No line parameters provided                                         \\
75Nerf \cite{75Nerf}         & 1 -- 10       & Lines are given in 72Nerf \cite{72Nerf}                                 \\
73JoMc   \cite{73JoMc}       & 1707 -- 1767  & No line parameters provided                                         \\
70TuThTo \cite{70TuThTo}     & 0.15 -- 0.15  & Line is given in 71TuToTh \cite{71TuToTh}                              \\
60OkHiSh \cite{60OkHiSh}     & 0.98          & Lines are given in 64OkTaMo \cite{64OkTaMo}                             \\
\hline\hline
\end{tabular}}
\end{table}

\section{The rovibrational database}

\subsection{Overview}
Tables~\ref{tab:marveldata} and \ref{tab:marvelnodata} present 
an overview of the experimental information considered during this project.
Each experimental source is given a tag composed of 
the last two digits of the year of publication and 
letters of the names of up to the first four co-authors.

Table~\ref{tab:marveldata} summarizes the sources included in our final 
MARVEL analysis and gives statistical information about the transitions 
these sources contain.
We utilized data from \nbSr\  literature sources,
including this work, yielding \nbTr\ transitions,
of which \nbNonRedTr\ are non-redundant.

Table \ref{tab:marvelnodata} lists sources that were \emph{not} included 
in the final analysis.
A number of other older sources \cite{60Oka,63ShKoTa,70KrGeShPo} were 
also excluded from the present analysis, as their measurements 
have been superseded by those of significantly more accurate studies. 

\subsection{Source-specific comments}

A significant problem we faced during the data collection was that 
there are several publications \cite{03PeKeFl,06FlLaSaSh,06PeVaDa} 
for \spec{h212c16o} where the authors did not provide the direct measured data
in their original paper and when approached they turned down our request 
to send the measured data forming the basis of their existing publication.
Similar problems, and related issues,
have been highlighted in a recent paper by Gordon \ea\ \cite{16GoPoBoEr}.
Seemingly it is not straightforward to include old data into new data compilations.

Another significant issue is that a number of sources 
do not provide a clear statement about the uncertainties of the observed lines.
Thus, we had to estimate them by various means,
which included comparisons with other sources as well as
combination-difference relations. 
Such sources include 76NaYaKu \cite{76NaYaKu},
82BrJoMcWo \cite{82BrJoMcWo}, 88ClVa \cite{88ClVa}, 
85TiChKuHu \cite{85TiChKuHu}, 77AlJoMc \cite{77AlJoMc}, 
77AlJoMcb \cite{77AlJoMcb}, and 10JaLaTcGa \cite{10JaLaTcGa}.
Since the authors of 10JaLaTcGa \cite{10JaLaTcGa} state 
that their line positions were not calibrated, 
we used an xMARVEL facility to calibrate the wavenumbers of the 
measured transitions published in 10JaLaTcGa. 
This gave a calibration factor of 1.000\,001\,148 for this source.

64OkTaMo \cite{64OkTaMo} provides high-resolution pure rotational transitions 
between levels in vibrationally excited states. 
However, 78DaWiBe \cite{78DaWiBe} found it necessary to reassign 
the vibrational states in this source;
we have adopted and in fact validated the assignments of 78DaWiBe.

76NaYaKu \cite{76NaYaKu} is an older and relatively 
low-resolution source, which nonetheless contains
lines for which no data are available from alternative sources.
The authors used a compact, non-standard notation
for the assignments, which had to be unpicked. 
A total of 134 lines could not be validated and thus were removed. 
Transitions involving high $K_a$ states did not resolve the $K$-doublets,
\ie, the $K_c$ quantum number.
For these transitions we used two lines corresponding to 
both possible transitions.

88NaReDaJo \cite{88NaReDaJo} gives only calculated line positions 
and residuals; thus, the observed frequencies were reconstructed
from this information.
Similarly, the tables in 79BrHuPi \cite{79BrHuPi} are of very poor quality 
and their line frequency data
could only be accurately reconstructed by extensive comparison 
with HITRAN \cite{17GoRoHiKo}; 88ClVa \cite{88ClVa}
also provides a small portion of 79BrHuPi's spectrum in readable form.

02BaCoHaPe \cite{02BaCoHaPe} recorded a spectrum of the 
$5\nu_2$ overtone; this spectrum was analyzed by
07SaBaHaRi \cite{07SaBaHaRi}, who also provided the data 
(C.M. Western, private communication, 2020). 
These studies did not contain an estimated uncertainty; thus,
an estimated value of 0.0015 \icm\
was adopted on the basis of combination difference relations.

72JoLoKi \cite{72JoLoKi} presents an extensive compilation of 
early microwave experiments on formaldehyde. 
Although this is a secondary source, 
some data were taken from here as the primary sources 
are not available to us.
72JoLoKi also contains tabulations of hyperfine-resolved 
transitions for the $1_{01} - 1_{11}$, $2_{11} - 2_{12}$,
and $3_{12} - 3_{13}$ rotational lines. 
Other microwave studies, including 59TaShSh \cite{59TaShSh}, 70TuThTo \cite{70TuThTo},
71TuToTh \cite{71TuToTh}, 72TuToTh \cite{72TuToTh}, and 17MuLe \cite{17MuLe} 
also present hyperfine-resolved data. 
Within the present study we completely neglect hyperfine effects and, 
where necessary, use central, hyperfine-unresolved line
positions.

\section{Results and Discussion}
\subsection{Validation}
According to Table~\ref{tab:marveldata}, 
we considered for validation \nbNonRedTr\  non-redundant transitions 
measured for \spec{h212c16o}.
13 of these transitions are artificial (see above) and 143 do not attach to the
principal components, they remain parts of floating components.
The transitions of the giant components of the SN of  \spec{h212c16o}
were validated using several techniques. 

193 transitions, including 
a few which did not obey the selection rules governing 
electric-dipole-allowed transitions, 
were removed at the first stage of validation.
In almost all cases these transitions were ones for which alternative, 
validated wavenumber entries were available from other sources.
The number of validated transitions within each segment 
is given in Table~\ref{tab:marveldata}. 
The  transitions which could not be validated are retained
in the transitions list given in the supplementary data, 
but they are given there as a negative wavenumber entry,
which means that they are ignored during the processing of the data by xMARVEL.
Therefore, these transitions do not contribute to the final empirical energy levels.

The other important step in the validation process involved comparisons with the 
AYTY line list \cite{15AlYuYaTe}.
These comparisons were performed iteratively between the empirical (xMARVEL)
and the AYTY energy levels.  
During the first phase,
the comparisons identified a number of issues with
the original xMARVEL dataset in the form of incorrect quantum 
numbers or scanning errors.
Once these were corrected, only a few lines were found for which the upper states
had no reasonable match within
the AYTY list. 
These transitions are augmented with a comment `theoretical mismatch' in 
the MARVEL XML file given in the supplementary material.
Those excluded lines which 
violate the electric-dipole selection rules have the comment `wrong labels' in the XML file.
At the end, we were able to validate \nbValTr\ of the collected transitions.

\begin{table}[!t]
\caption{MARVEL-based vibrational band origins (VBO) of \spec{h212c16o} and their 
        literature counterparts obtained from accurate effective Hamiltonian fits.$^a$
}
\label{tab:vibterms}
\small
\vspace{-0.5cm}
\begin{center}
\begin{tabular}{lcrrrrr}
\hline \hline 
     Band           & Symmetry & VBO(MARVEL)       & $N_0$ & $N_{\rm RL}$ & $J_{\rm max}$ & VBO(lit)                           \\ \hline
     $\nu_0$        & a$_1$    & 0.0                & 31    & 525          & 38            & 0.0                                \\
     $\nu_4$        & b$_1$    & 1167.256\,76(10)   & 6     & 346          & 30            & 1167.258(2) \cite{77AlJoMc}        \\
                    &          &                    &       &              &               & 1167.256\,28(2) \cite{03PeKeFl}    \\
     $\nu_6$        & b$_2$    & 1249.094\,42(10)   & 3     & 405          & 30            & 1249.091(2) \cite{77AlJoMc}        \\
                    &          &                    &       &              &               & 1249.094\,68(2) \cite{03PeKeFl}    \\
     $\nu_3$        & a$_1$    & 1500.174\,50(10)   & 4     & 382          & 30            & 1500.176(3) \cite{77AlJoMc}        \\
                    &          &                    &       &              &               & 1500.174\,74(12) \cite{03PeKeFl}   \\
     $\nu_2$        & a$_1$    & 1746.009\,14(10)   & 4     & 446          & 37            & 1746.008\,9(1) \cite{82BrJoMcWo}   \\ 
                    &          &                    &       &              &               & 1746.008\,86(13) \cite{03PeKeFl}   \\
     $2\nu_4$       & a$_1$    & --                 & 0     & 65           & 11            & 2327.523\,9(5) \cite{06PeVaDa}     \\ 
                    &          &                    &       &              &               & 2327.1(8) \cite{79HaTi}            \\
     $\nu_4+\nu_6$  & a$_2$    & --                 & 0     & 47           & 11            & 2422.970\,1(50) \cite{06PeVaDa}    \\
     $2\nu_6$       & a$_1$    & --                 & 0     & 81           & 25            & 2494.354\,3(5) \cite{06PeVaDa}     \\
     $\nu_3+\nu_4$  & b$_1$    & --                 & 0     & 15           & 10            & 2667.048\,1(20) \cite{06PeVaDa}    \\ 
                    &          &                    &       &              &               & 2655.5 \cite{79BrHuPi}             \\
     $\nu_3+\nu_6$  & b$_2$    & 2719.156\,04(50)   & 1     & 232          & 25            & 2719.155\,9(10) \cite{79BrHuPi}    \\
     $\nu_1$        & a$_1$    & 2782.456\,91(50)   & 2     & 440          & 36            & 2782.456\,9(10) \cite{79BrHuPi}    \\
     $\nu_5$        & b$_2$    & 2843.323\,54(12)   & 4     & 388          & 35            & 2843.325\,6(10) \cite{79BrHuPi}    \\
     $\nu_2+\nu_4$  & b$_1$    & --                 & 0     & 153          & 24            & 2905.968\,5(20) \cite{06PeVaDa}    \\ 
                    &          &                    &       &              &               & 2905(1) \cite{79BrHuPi}            \\
     $2\nu_3$       & a$_1$    & --                 & 0     & 38           & 25            & 2998.987\,3(5) \cite{06PeVaDa}     \\ 
                    &          &                    &       &              &               & 2999.5(5) \cite{79BrHuPi}          \\
     $\nu_2+\nu_6$  & b$_2$    & 3000.065\,74(50)   & 1     & 100          & 21            & 3000.065\,6(10) \cite{79BrHuPi}    \\
     $2\nu_2$       & a$_1$    & 3471.720\,8(78)    & 2     & 373          & 31            & 3471.718(4)  \cite{73Toth}         \\
     $\nu_1+\nu_4$  & b$_1$    & 3941.530\,8(10)    & 1     & 30           & 10            & --                                 \\
     $\nu_4+\nu_5$  & b$_2$    & 3996.518\,6(10)    & 1     & 35           & 9             & --                                 \\
     $\nu_3+\nu_5$  & b$_2$    & 4335.0989\,4(10)   & 1     & 170          & 24            & --                                 \\
     $2\nu_2+\nu_6$ & b$_2$    & --                 & 0     & 207          & 26            & 4734.207\,8(50) \cite{06FlLaSaSh}  \\
     $3\nu_2$       & a$_1$    & --                 & 0     & 176          & 26            & 5177.759\,52(70) \cite{06FlLaSaSh} \\
     $2\nu_5$       & a$_1$    & 5651.196\,4(15)    & 1     & 179          & 20            & --                                 \\
\hline\hline
\end{tabular}
\end{center}
\vspace{-0.2cm}
$^a$The VBO values are in \icm.
The uncertainties of the last VBO digits are given in parentheses. 
The symmetry of the band, the number of transitions determining a particular VBO ($N_0$), 
as well as the number of rotational levels ($N_{\rm RL}$) and the maximum $J$ value 
within a given vibrational band are also displayed.
\end{table}


\subsection{Final energies}
The MARVEL analysis yielded, in total, \nbEl\  validated empirical rovibrational 
energies for \spec{h212c16o}; they are provided in the supplementary material.
The highest rotational quantum number is $J_{\rm max} = 38$ 
and the empirical energies go up to 
6188 \icm; there are transitions which probe energy
levels higher than this but none of them are assigned.
Thus, all empirical energy levels are well below the 
minima corresponding to hydroxymethylene.
A large number of further experimental studies are needed to reach that dynamically
important and interesting region.
The region covered is also considerably more limited than that covered 
by SEP measurements in the 1980s \cite{84ReFiKiDa}.

Table~\ref{tab:vibterms} presents the vibrational band origins (VBO) 
and a summary of the number of empirical rovibrational energy levels 
determined for each 
vibrational state. 
The bands $2\nu_3+\nu_6$, $\nu_2+\nu_5$, and $\nu_2+\nu_3+\nu_4$, 
which have a combined total of only 6 levels, have been omitted from the table. 
Table~\ref{tab:vibterms} also gives term values of this and previous studies, 
\ie, levels with $J=0$ for given vibrational bands.

Table~\ref{tab:vibterms} contains term values for 14 vibrational band origins,
including the six fundamentals, 
determined during this study.
The term values of the $\nu_2$, $\nu_3$, and $\nu_4$ fundamentals are 
determined particularly accurately, in part because in these cases the 
upper states have been subjected to high-accuracy microwave studies;
indeed, such studies for $\nu_3$ and $\nu_4$ form part of the present work.
We note that these fundamentals are all characterized 
by at least three transitions. 
In contrast, the term values of the combination bands 
are much less accurate and determined by only one, 
or in one case two, transitions. 
It is important to note that the fundamentals of Ref. \cite{77AlJoMc}
result from effective-Hamiltonian calculations,
which may yield higher apparent accuracy than the present MARVEL treatment.

\begin{figure}[t!]
\centering
	\includegraphics[width=\linewidth]{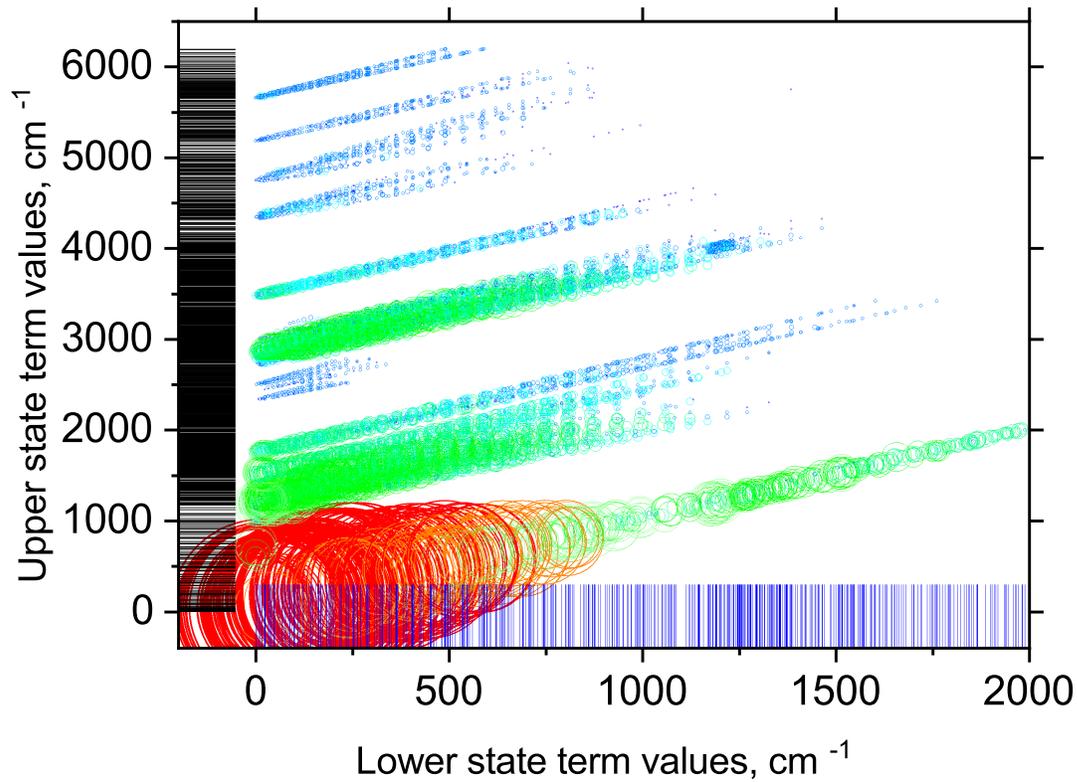}
	\caption{The upper-state energies of the experimental  transitions used in this work, 
	against corresponding lower-state energies of \spec{h212c16o}. 
	The vertical bars along the horizontal axis show the lower state energies, 
	while the horizontal bars along the vertical axis give the upper state energies.  
	Each circle represents a particular transition, with the size proportional 
	to the number of transitions supporting the corresponding upper state. 
	This value ranges from 1 (dark blue) to 102 (red).}
	\label{fig:H2CO_CD}
\end{figure}

Figure~\ref{fig:H2CO_CD} provides a visual representation of the 
spectroscopic network of \spec{h212c16o}, showing the upper-state energies
of the measured transitions against corresponding lower-state energies,
and giving a visual impression of the number of transitions linking them.

\begin{table}[t!]
\caption{Pure rotational frequencies of \spec{h212c16o}\ lines under protection
by the National Academies of Sciences, Engineering, and Medicine (NASEM) \cite{15National} and their various experimental and empirical determinations.
The uncertainties of the last few frequency digits
are indicated in parentheses.} \label{tab:nasem} 
\small
\vspace{-0.5cm}
\begin{center}
\begin{tabular}{lcrl} 
\hline \hline 
$f_{\rm NASEM}$/GHz & Line assignment                            & $f_{\rm MARVEL}$/GHz  & $f_{\rm expt}$/GHz                    \\ \hline
4.829\,66           & $\nu_0, 1_{1,0} \leftarrow 1_{1,1}$ & 4.829\,659\,960(50)   & 4.829\,659\,960(50) \cite{71TuToTh}   \\
                    &                                            &                       & 4.829\,660\,0(10) \cite{96BoDePoLi}   \\    
                    &                                            &                       & 4.829\,649\,0(50) \cite{73ChGu}       \\    
                    &                                            &                       & 4.829\,660(10) \cite{72JoLoKi}        \\
14.488              & $\nu_0, 2_{1,1} \leftarrow 2_{1,2}$ & 14.488\,478\,810(80)  & 14.488\,478\,810(80) \cite{72TuToTh}  \\
                    &                                            &                       & 14.488\,479\,0(10) \cite{96BoDePoLi}  \\    
                    &                                            &                       & 14.488\,650(20) \cite{64OkTaMo}       \\  
                    &                                            &                       & 14.488\,650(50) \cite{51LaSt}         \\    
                    &                                            &                       & 14.488\,65(10) \cite{73ChGu}          \\
140.84              & $\nu_0, 2_{1,2} \leftarrow 1_{1,1}$ & 140.839\,502(10)      & 140.839\,502(10) \cite{96BoDePoLi}    \\
                    &                                            &                       & 140.839\,526(32) \cite{72Nerf}        \\    
                    &                                            &                       & 140.839\,300(50) \cite{56Erlandss}    \\  
145.603             & $\nu_0, 2_{02} \leftarrow 1_{0,1}$  & 145.602\,949(10)      & 145.602\,949(10) \cite{96BoDePoLi}    \\
                    &                                            &                       & 145.602\,966(34) \cite{72Nerf}        \\    
                    &                                            &                       & 145.603\,100(50) \cite{56Erlandss}    \\  
150.498             & $\nu_0, 2_{1,1} \leftarrow 1_{1,0}$ & 150.498\,334(10)      & 150.498\,334(10) \cite{96BoDePoLi}    \\
                    &                                            &                       & 150.498\,355(34) \cite{72Nerf}        \\    
                    &                                            &                       & 150.498\,200(50) \cite{56Erlandss}    \\  
218.222             & $\nu_0, 3_{0,3} \leftarrow 2_{0,3}$ & 218.222\,192(10)      & 218.222\,192(10) \cite{96BoDePoLi}    \\
                    &                                            &                       & 218.222\,186(49) \cite{72Nerf}        \\    
                    &                                            &                       & 218.221\,600(50) \cite{56Erlandss}    \\ \hline\hline
\end{tabular}
\end{center}
\end{table}

\subsection{Protected formaldehyde lines}
The National Academies of Sciences, Engineering, and Medicine (NASEM)
maintains a list of key long-wavelength transitions relevant for
astrophysical studies \cite{15National}. 
This list contains six transitions belonging to  
\spec{h212c16o}; they are given in Table~\ref{tab:nasem}. 
Some previous MARVEL studies, notably on water \cite{19ToFuTeCs,20FuToTePo} 
and ammonia \cite{20FuCoTeYu}, allowed the accuracy, 
to which some of these lines were known, to be improved using MARVEL.
In the case of \spec{h212c16o}, however, 
we find that the uncertainties in our  MARVEL-determined frequencies 
simply match those of the current best laboratory determinations; 
for a discussion how these uncertainties are determined, see Ref.~\cite{19ToFuTeCs}.
Nevertheless, Table~\ref{tab:nasem} is still useful as it shows all
highly-accurate studies of the NASEM-protected lines.

\begin{table}[!ht]
\caption{Extract from the H$_2$CO state file. The full table is available from \url{www.exomol.com}.$^a$} \label{tab:states}
\begin{center}
\footnotesize
\tabcolsep=1pt
\vspace{-0.5cm}
\resizebox{0.9\columnwidth}{!}{%
\renewcommand{\arraystretch}{1.0}
\begin{tabular}{rR{2.0cm}R{0.3cm}R{0.3cm}R{1.4cm}R{1.7cm}C{0.8cm}ccccccC{0.8cm}ccC{1.0cm}ccccccC{1.0cm}r}
\hline \hline
        $I$  &  $\tilde{E}$/\icm   &  $g$  &  $J$ & $\delta$/\icm & $\tau$/s&  $\mathit{\Gamma}_{\rm tot}$ & $v_1$ & $v_2$ & $v_3$ & $v_4$ & $v_5$ & $v_6$ & $\mathit{\Gamma}_{\rm vib}$  & $K_a$  & $K_c$ & $|C_i^{2}|$ & $n_1$ & $n_2$ & $n_3$ & $n_4$ & $n_5$ & $n_6$ & K  &$\tilde{E}_{\rm AYTY}$/\icm  \\
\hline
    1 &      0.000000 &   1 &    0 &    0.000001 &  Inf        &  A1  &   0 &   0 &   0 &   0 &   0 &   0 &  A1  &   0 &   0 &  1.00 &   0 &   0 &   0 &   0 &   0 &   0 &   Ma   &    0.000000   \\
    2 &   1500.174503 &   1 &    0 &    0.000100 &  3.1423E-01 &  A1  &   0 &   0 &   1 &   0 &   0 &   0 &  A1  &   0 &   0 &  1.00 &   0 &   0 &   0 &   0 &   1 &   0 &   Ma   &   1500.120955 \\
    3 &   1746.009136 &   1 &    0 &    0.000100 &  3.1935E-02 &  A1  &   0 &   1 &   0 &   0 &   0 &   0 &  A1  &   0 &   0 &  1.00 &   1 &   0 &   0 &   0 &   0 &   0 &   Ma   &   1746.045388 \\
    4 &   2327.497142 &   1 &    0 &    0.200000 &  4.7496E-01 &  A1  &   0 &   0 &   0 &   2 &   0 &   0 &  A1  &   0 &   0 &  1.00 &   0 &   0 &   0 &   0 &   0 &   2 &   Ca   &   2327.497142 \\
    5 &   2494.322937 &   1 &    0 &    0.200000 &  2.0931E-01 &  A1  &   0 &   0 &   0 &   0 &   0 &   2 &  A1  &   0 &   0 &  1.00 &   0 &   0 &   0 &   1 &   1 &   0 &   Ca   &   2494.322937 \\
    6 &   2782.456913 &   1 &    0 &    0.000500 &  1.5673E-02 &  A1  &   1 &   0 &   0 &   0 &   0 &   0 &  A1  &   0 &   0 &  1.00 &   0 &   0 &   1 &   0 &   0 &   0 &   Ma   &   2782.410921 \\
    7 &   2999.006647 &   1 &    0 &    0.200000 &  1.0513E-01 &  A1  &   0 &   0 &   0 &   0 &   0 &   2 &  A1  &   0 &   0 &  1.00 &   0 &   0 &   0 &   1 &   1 &   0 &   Ca   &   2999.006647 \\
    8 &   3238.937891 &   1 &    0 &    0.200000 &  2.9533E-02 &  A1  &   0 &   1 &   1 &   0 &   0 &   0 &  A1  &   0 &   0 &  1.00 &   1 &   0 &   0 &   0 &   1 &   0 &   Ca   &   3238.937891 \\
    9 &   3471.720843 &   1 &    0 &    0.007770 &  1.5448E-02 &  A1  &   0 &   2 &   0 &   0 &   0 &   0 &  A1  &   0 &   0 &  1.00 &   2 &   0 &   0 &   0 &   0 &   0 &   Ma   &   3471.719306 \\
   10 &   3825.967015 &   1 &    0 &    0.300000 &  1.6623E-01 &  A1  &   0 &   0 &   1 &   2 &   0 &   0 &  A1  &   0 &   0 &  1.00 &   0 &   0 &   0 &   0 &   1 &   2 &   Ca   &   3825.967015 \\
   11 &   3936.435541 &   1 &    0 &    0.300000 &  3.3267E-02 &  A1  &   0 &   0 &   3 &   0 &   0 &   0 &  A1  &   0 &   0 &  1.00 &   0 &   0 &   0 &   3 &   0 &   0 &   Ca   &   3936.435541 \\
   12 &   4058.101422 &   1 &    0 &    0.300000 &  3.0046E-02 &  A1  &   0 &   1 &   0 &   2 &   0 &   0 &  A1  &   0 &   0 &  1.00 &   1 &   0 &   0 &   0 &   0 &   2 &   Ca   &   4058.101422 \\

\hline
\hline
\end{tabular}}
\renewcommand{\arraystretch}{0.52}
\end{center}
\noindent
$^a$
$I$: state identifier; 
$\tilde{E}$: state term value; 
$g$: state degeneracy; 
$J$: state rotational quantum number; 
$\delta$: energy uncertainty;
$\tau$: lifetime; 
$\mathit{\Gamma}_{\rm tot}$: total symmetry in $C_{2\nu}(M)$; 
$v_1 - v_6$: normal mode vibrational quantum numbers; 
$\mathit{\Gamma}_{\rm vib}$: symmetry of vibrational contribution in $C_{2\nu}(M)$; 
$K_a$ and $K_c$: rotational quantum numbers; 
$|C_i^{2}|$: largest coefficient used in the assignment; 
$n_1 - n_6$: TROVE vibrational quantum numbers; 
K: data kind indicating if the term value is based on the MARVEL (`Ma') 
or the AYTY energy list (`Ca'); 
$\tilde{E}_{\rm AYTY}$: original AYTY state term value. \\
\end{table}

\section{Updated AYTY line list}

The AYTY line list \cite{15AlYuYaTe} of \spec{h212c16o} contains approximately 
10 billion transitions linking 10.3 million rotational-vibrational states.
These states are those with $J\leq 70$ which lie up to 10~000 \icm\ above 
the ground state.
As part of this study, we replace those energy levels of the AYTY list
which are determined empirically by the MARVEL process. 

The rovibrational states were matched on the basis of quantum numbers 
and then checked using the energies. 
As a result, 367\,779 transition frequencies are determined
using the empirical energy levels of this study.
Of these,  183\,673 lie above the dynamic HITRAN intensity cutoff \citep{13DoHiYuTe}. 
These numbers should be compared to the \nbNonRedTr\ non-redundant transitions 
which form the input to the original spectroscopic network of \spec{h212c16o}. 
It can be seen that this process leads to a significant increase 
in the number of transitions
whose line centres are determined to high-resolution experimental accuracy.

Obviously, \nbEl\ empirically determined energy levels is only 
a small portion of the 10.3 million entries in the original AYTY line list.
We therefore had to estimate the (significantly larger) uncertainties 
associated with the calculated energy levels. 
This is important, not least because it allows transitions
that are predicted with high accuracy to be easily identified.
We estimated the uncertainties, in \icm, of the computed levels using the formula 
$$
\delta = 0.1\, (v_1 + v_2 + v_3 + v_4 +v_5+v_6) + 0.005 J(J+1).
$$
This update is in line with the requirements of the 2020 
release of ExoMol \cite{20LaScTe} and is designed to facilitate 
the use of the data in high-resolution studies by quantifying
the uncertainty associated with each transition and 
hence identifying those known with the required accuracy
for a given study.

\begin{figure}[t!]
\centering
\includegraphics[width=0.312\linewidth]{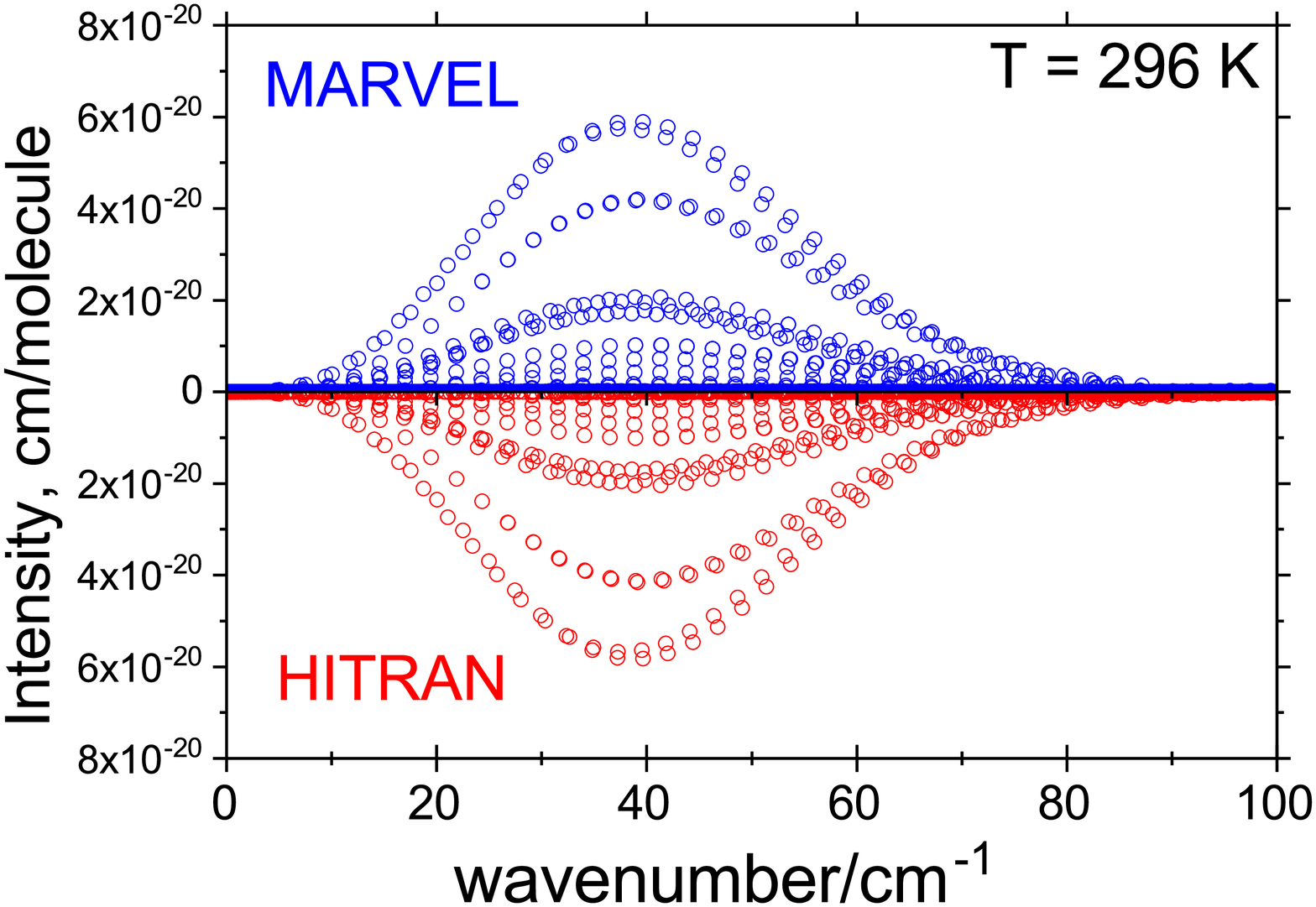}
\includegraphics[width=0.312\linewidth]{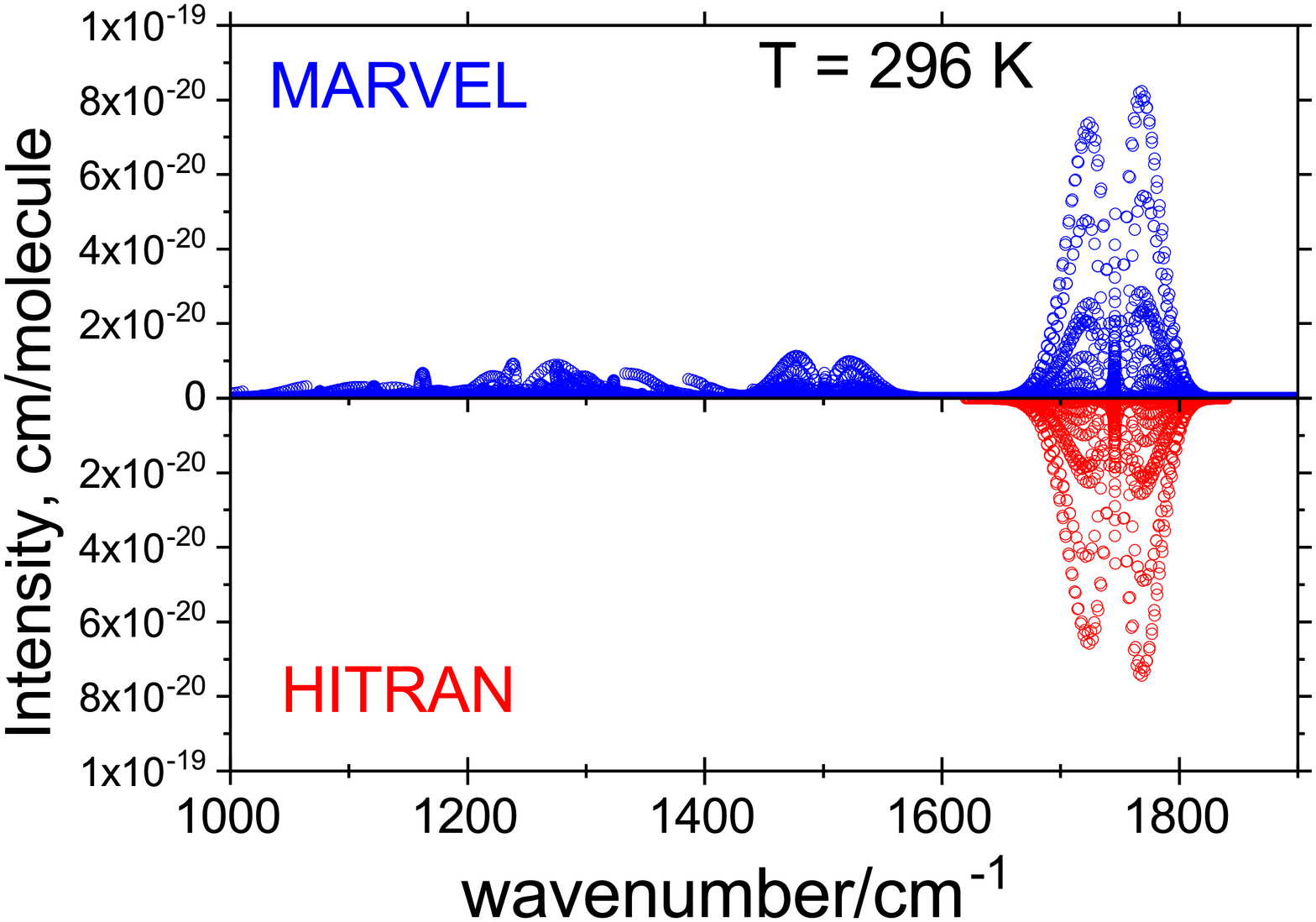}
\includegraphics[width=0.312\linewidth]{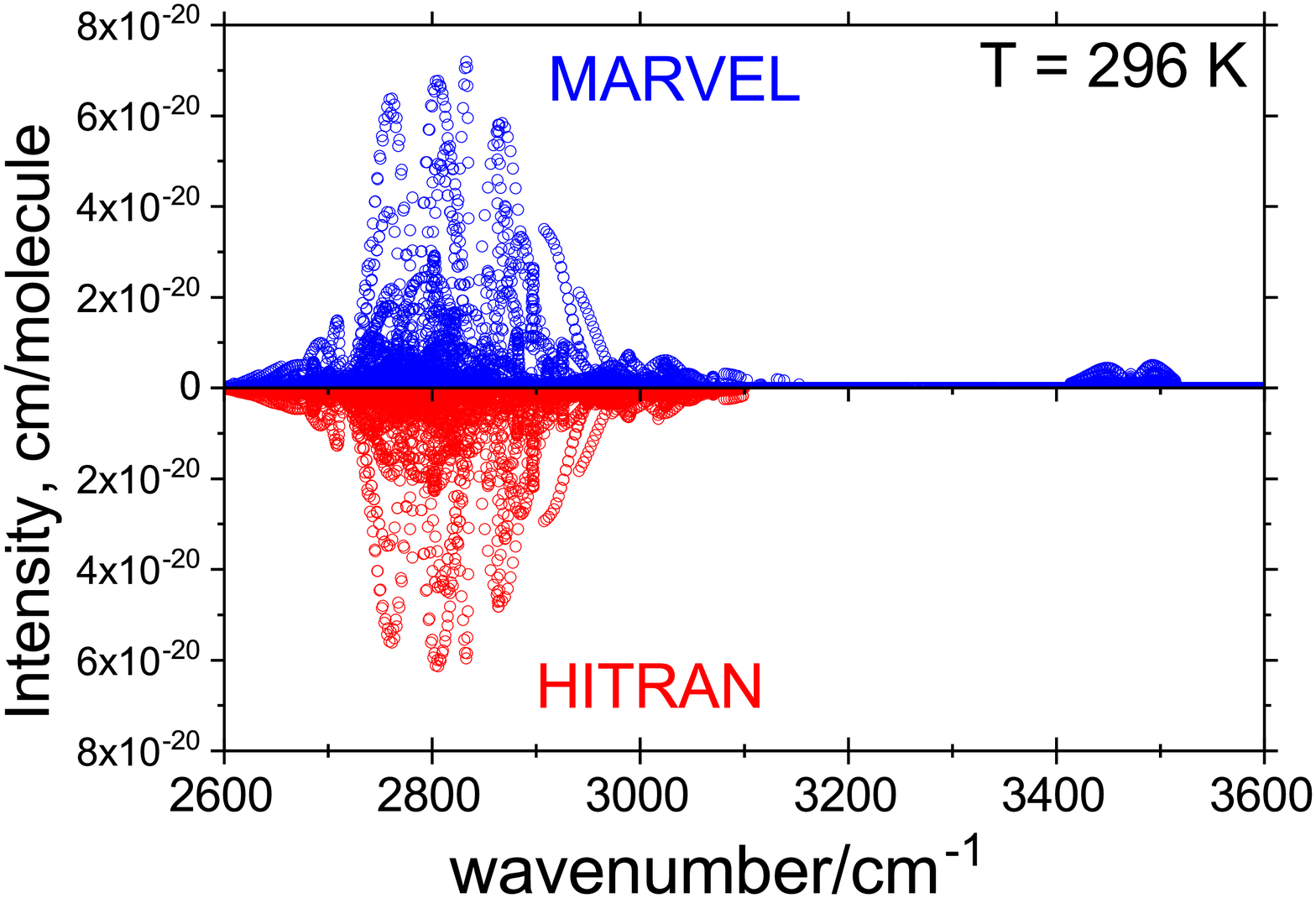}
	\caption{Room-temperature ($T=296$~K) spectra of \spec{h212c16o}
	in three different regions covered by HITRAN\,2016 \cite{17GoRoHiKo}. 
	The upper panels, in blue, show stick spectra simulated using the MARVEL 
	energy term values from this work and the Einstein-$A$ coefficients
	from the AYTY line. 
	The lower panels, in red, show the corresponding spectra taken 
	from HITRAN\,2016. 
	}
	\label{fig:H2CO-MARVEL-vs-HITRAN}
\end{figure}

Table~\ref{tab:states} gives a small portion of the updated ExoMol States file
in standard ExoMol format \cite{13TeHiYu,20LaScTe}. 
The complete file along with the AYTY Trans file, 
which is unchanged by the current procedure, can be obtained from the 
ExoMol website (\url{www.exomol.com}).

\begin{figure}[t!]
\centering
		\includegraphics[width=0.75\linewidth]{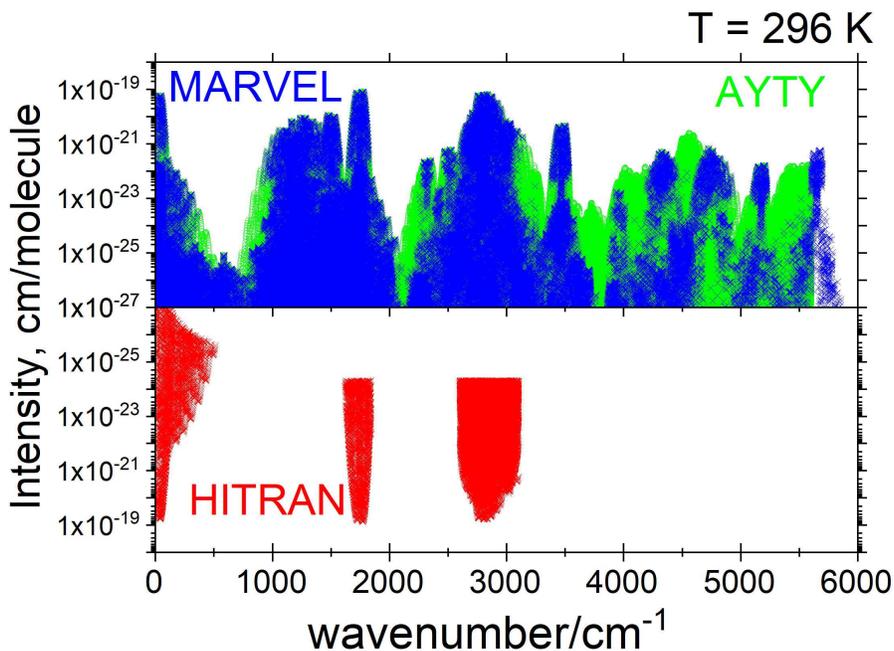}

\caption{Room temperature ($T=296$~K) spectra of \spec{h212c16o} 
from three different sources: lower panel, HITRAN\,2016~\cite{17GoRoHiKo};
upper panel, this work (see caption of Fig.~\protect\ref{fig:H2CO-MARVEL-vs-HITRAN}) where possible or else from the AYTY~\protect\cite{15AlYuYaTe} line list.}
	\label{fig:H2CO-MARVEL-vs-HITRAN-vs-TROVE}
\end{figure}

The HITRAN database \cite{17GoRoHiKo} contains limited data on \spec{h212c16o}:
besides pure rotations it covers only two bands.
Figure~\ref{fig:H2CO-MARVEL-vs-HITRAN} gives a comparison for these regions.
The agreement is very good and shows that our MARVEL analysis 
is largely sufficient to cover these regions.
Figure~\ref{fig:H2CO-MARVEL-vs-HITRAN-vs-TROVE} illustrates 
the relative (in)completeness of the HITRAN and our MARVEL data sets 
when used to simulate the room temperature spectra of \spec{h212c16o}. 
There are a number of missing bands in HITRAN and
the database can now be supplemented with our synthetic line list 
constructed using the line positions determined to experimental
accuracy using MARVEL and AYTY transition intensities.

\begin{figure}[t!]
\centering
\includegraphics[width=0.75\linewidth]{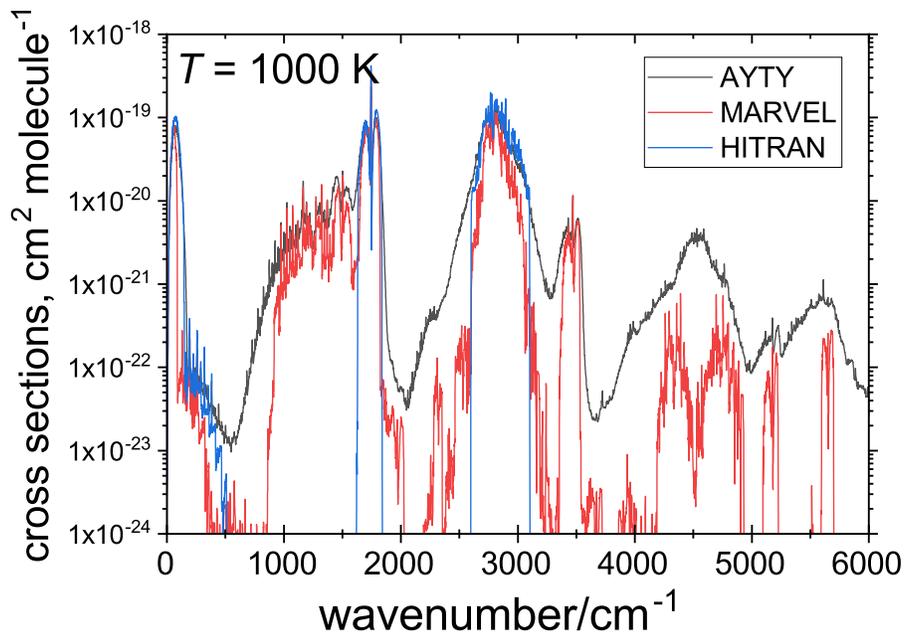}
\caption{High temperature ($T=1000$~K) spectra of \spec{h212c16o}
from three different sources: HITRAN\,2016~\cite{17GoRoHiKo}, 
AYTY~\cite{15AlYuYaTe}, and this work 
(see also caption to Fig.~\ref{fig:H2CO-MARVEL-vs-HITRAN}).  }
	\label{fig:H2CO-MARVEL-vs-HITRAN-1000K}
\end{figure}

The AYTY line list was actually designed for treating formaldehyde
in hot environments. 
In order to illustrate the lack  of experimental data at high temperatures, Fig.~\ref{fig:H2CO-MARVEL-vs-HITRAN-1000K} shows  $T = $ 1000~K 
absorption spectra simulated with the same three source as above: 
MARVEL line positions (this work) combined with the 
AYTY Einstein-$A$ coefficients, the HITRAN line list, and the AYTY line list.
It is clear that at high temperatures only AYTY provides
the spectroscopic coverage required for high temperatures and that known laboratory spectra 
are not capable of giving a complete representation of hot formaldehyde.

\section{Summary and Conclusions}
Apart from a few sources where the authors were reluctant to provide
their published results for analysis \cite{03PeKeFl,06FlLaSaSh,06PeVaDa},
all literature lines measured for \spec{h212c16o}\  have been collected and
analyzed during this study.
The information contained in the literature sources 
was augmented by new measurements results, including seven Lamb-dip lines
and 82 further lines, all corresponding to rotational transitions
within the ground, $\nu_3$, $\nu_4$, and $\nu_6$ vibrational states.
The analysis utilized the Measured Active Rotational-Vibration Levels (MARVEL)
approach and the xMARVEL code. 
A characteristic of MARVEL is that it is an ``active'' approach,
which means that should new sources, or even existing sources 
that have not been included, became available it is straightforward 
to include them in a renewed analysis and produce 
an updated version of the line-by-line database assembled for \spec{h212c16o}.

Of the total of \nbTr\ transitions processed, 
which come from \nbSr\ sources including this work,
\nbValTr\   could be validated, providing \nbEl\  empirical energy
levels of \spec{h212c16o}\  with statistically well-defined uncertainties. 
Our newly-determined empirical rotational-vibrational energy levels are used to 
improve the accuracy of ExoMol's AYTY line list for hot formaldehyde \cite{15AlYuYaTe}. 
This improved line list is available as supplementary material,
along with the xMARVEL 
files, containing all the rovibrational
transitions collated, whether validated or not, 
as well as all the empirical energy levels derived during this study.

It is the stated plan of the ExoMol project \cite{12TeYu} 
to update all ExoMol line lists to include both empirical energy levels, 
where available, and uncertainties for all the energy levels; 
thus allowing the uncertainty associated with any transition to be estimated. 
These data will become a standard part of the ExoMol line list and
also form input for the new ExoMolHR database, which provides data only on lines of 
higher intensity whose wavelengths are known to high accuracy.
So far only a limited number of ExoMol line lists are available  with
specified uncertainties in the energy levels. 
These include important line lists for which MARVEL data were already available, 
namely those for water \cite{18PoKyZoTe}, AlH \cite{18YuWiLeLo}, 
\spec{12c2} \cite{20McSyBoYu}, \spec{12c2h2} \cite{19McMaHoPe} and \spec{48ti16o} \cite{19McMaHoPe},
and recently produced line lists including those for
\spec{co2} \cite{20YuMeFrTe}, \spec{h3o+} \cite{20YuTeMiMe} and \spec{nh3} \cite{19CoYuTe}. 
The present study is the first in a series where a MARVEL analysis is
performed with the intention of updating an available line list.

\section*{Acknowledgments}
We thank Dr. Colin Western for providing the data generated 
by 07SaBaHaRi \cite{07SaBaHaRi}.  
The work at UCL is supported by STFC Projects No. ST/M001334/1 
and ST/R000476/1, and the European
Research Council (ERC) under the European Union’s Horizon 2020 
research and innovation programme through Advance Grant number  883830.
The work performed in Budapest received support from NKFIH 
(grant no. K119658), from the grant VEKOP-2.3.2-16-2017-000, and
from the  ELTE Institutional Excellence Program (TKP2020-IKA-05) 
financed by the Hungarian Ministry of Human Capacities.
The work performed in Bologna was supported by the Universit\`a di Bologna 
(RFO funds) and by MIUR (Project PRIN 2015: STARS in the CAOS, 
Grant Number 2015F59J3R).
\setstretch{1.05}
\setlength{\bibsep}{1.0pt plus 0.0ex}
\bibliographystyle{elsarticle-num}

\begin{thebibliography}{100}
\expandafter\ifx\csname url\endcsname\relax
  \def\url#1{\texttt{#1}}\fi
\expandafter\ifx\csname urlprefix\endcsname\relax\def\urlprefix{URL }\fi
\expandafter\ifx\csname href\endcsname\relax
  \def\href#1#2{#2} \def\path#1{#1}\fi

\bibitem{94Mclaughl}
J.~K. McLaughlin, {Formaldehyde and cancer: a critical review}, Int. Arch. Occ.
  Env. Hea. 66 (1994) 295--301.

\bibitem{10NiWo}
G.~D. Nielsen, P.~Wolkoff, {Cancer effects of formaldehyde: a proposal for an
  indoor air guideline value}, Arch. Toxicol. 84 (2010) 423--446.

\bibitem{20HeLiFe}
Q.~He, J.~Li, Q.~Feng, {Ppb-level formaldehyde detection system based on a 3.6
  $\mu$m interband cascade laser and mode-locked cavity enhanced absorption
  spectroscopy with self-calibration of the locking frequency}, Infrared Phys.
  Techn. 105 (2020) 103205.

\bibitem{75MoWa}
D.~C. Moule, A.~D. Walsh, {Ultraviolet spectra and excited states of
  formaldehyde}, Chem. Rev. 75 (1975) 67--84.

\bibitem{83ClRa}
D.~J. Clouthier, D.~A. Ramsay, {The spectroscopy of formaldehyde and
  thioformaldehyde}, Annu. Rev. Phys. Chem. 34 (1983) 31--58.

\bibitem{83MoWe}
C.~B. Moore, J.~C. Weisshaar, {Formaldehyde photochemistry}, Annu. Rev. Phys.
  Chem. 34 (1983) 525--555.

\bibitem{00Wayne}
R.~P. Wayne, {Chemistry of atmospheres}, Oxford University Press, New York,
  2000.

\bibitem{04ToLaLeCh}
D.~Townsend, S.~A. Lahankar, S.~K. Lee, S.~D. Chambreau, A.~G. Suits, X.~Zhang,
  J.~Rheinecker, L.~B. Harding, J.~M. Bowman, The roaming atom: straying from
  the reaction path in formaldehyde decomposition, Science 306 (2004)
  1158--–1161.

\bibitem{09ZhMaMoBr}
P.~Zhang, S.~Maeda, K.~Morokuma, B.~J. Braams, {Photochemical reactions of the
  low-lying excited states of formaldehyde: $T_1/S_0$ intersystem crossings,
  characteristics of the $S_1$ and $T_1$ potential energy surfaces, and a
  global $T_1$ potential energy surface}, J. Chem. Phys. 130 (2009) 114304.

\bibitem{84ReFiKiDa}
D.~E. Reisner, R.~W. Field, J.~L. Kinsey, H.-L. Dai, {Stimulated emission
  spectroscopy: a complete set of vibrational constants for $\tilde{\rm
  X}\,^{1}A_{1}$ formaldehyde}, J. Chem. Phys. 80 (1984) 5968--5978.

\bibitem{11Csaszar}
A.~G. Cs\'asz\'ar, {Anharmonic molecular force fields}, WIREs Comput. Mol. Sci.
  2 (2012) 273--289.

\bibitem{93MaLeTa}
J.~M.~L. Martin, T.~J. Lee, P.~R. Taylor, {An accurate \emph{ab initio} quartic
  force-field for formaldehyde and its isotopomers}, J. Mol. Spectrosc. 160
  (1993) 105--116.

\bibitem{95CaPiHa}
S.~Carter, N.~Pinnavaia, N.~C. Handy, {The vibrations of formaldehyde}, Chem.
  Phys. Lett. 240 (1995) 400--408.

\bibitem{96BuMcSi}
D.~C. Burleigh, A.~B. McCoy, E.~L. Sibert, {An accurate quartic force field for
  formaldehyde}, J. Chem. Phys. 104 (1996) 480--487.

\bibitem{18MoMaRiAg}
W.~J. Morgan, D.~A. Matthews, M.~Ringholm, J.~Agarwal, J.~Z. Gong, K.~Ruud,
  W.~D. Allen, J.~F. Stanton, H.~F. Schaefer, {Geometric energy derivatives at
  the complete basis set limit: application to the equilibrium structure and
  molecular force field of formaldehyde}, J. Chem. Phys. 14 (2018) 1333--1350.

\bibitem{82JeBu}
P.~Jensen, P.~Bunker, {The geometry and the inversion potential function of
  formaldehyde in the A\,$^1A_2$ and a$^3A_2$ electronic states}, J. Mol.
  Spectrosc. 94 (1982) 114--125.

\bibitem{81GoYaSc}
J.~D. Goddard, Y.~Yamaguchi, H.~F. Schaefer, {Features of the H$_2$CO potential
  energy hypersurface pertinent to formaldehyde photodissociation}, J. Chem.
  Phys. 75 (1981) 3459--3465.

\bibitem{03JaCh}
A.~F. Jalbout, C.~M. Chang, {The H$_2$CO potential energy surface: advanced
  \emph{ab initio} and density functional theory study}, J. Mol. Struc.
  (THEOCHEM) 634 (2003) 127--135.

\bibitem{04ZhZoHaBo}
X.~B. Zhang, S.~L. Zou, L.~B. Harding, J.~M. Bowman, A global \emph{ab initio}
  potential energy surface for formaldehyde, J. Phys. Chem. A 108 (2004)
  8980--8986.

\bibitem{08KoWaBrBo}
L.~Koziol, Y.~Wang, B.~J. Braams, J.~M. Bowman, A.~I. Krylov, {The theoretical
  prediction of infrared spectra of trans- and cis-hydroxycarbene calculated
  using full dimensional \emph{ab initio} potential energy and dipole moment
  surfaces}, J. Chem. Phys. 128 (2008) 204310.

\bibitem{09UlBeLeGr}
O.~N. Ulenikov, E.~S. Bekhtereva, C.~Leroy, O.~V. Gromova, A.~L. Fomchenko, {On
  the determination of the intramolecular potential energy surface of
  polyatomic molecules: hydrogen sulfide and formaldehyde as an illustration},
  J. Mol. Spectrosc. 255 (2009) 88--100.

\bibitem{11YaYuJeTh}
A.~Yachmenev, S.~N. Yurchenko, P.~Jensen, W.~Thiel, {A new ``spectroscopic''
  potential energy surface for formaldehyde in its ground electronic state}, J.
  Chem. Phys. 134 (2011) 244307.

\bibitem{08ScRePiSi}
P.~R. Schreiner, H.~P. Reisenauer, F.~C. Pickard, A.~C. Simmonett, W.~D. Allen,
  E.~M\'atyus, A.~G. Cs\'asz\'ar, {Capture of hydroxymethylene and its fast
  disappearance through tunnelling}, Nature 453 (2008) 906--909.

\bibitem{12JaMcSz}
P.~Jankowski, A.~McKellar, K.~Szalewicz, {Theory untangles the high-resolution
  infrared spectrum of the \emph{ortho}-H$_2$CO van der Waals complex}, Science
  336 (2012) 1147--1150.

\bibitem{17PaSzCs}
D.~Papp, T.~Szidarovszky, A.~G. Cs\'asz\'ar, {A general variational approach
  for computing rovibrational resonances of polyatomic molecules. Application
  to the weakly bound H$_2$He$^+$ and H$_2\cdot$CO systems}, J. Chem. Phys. 147
  (2017) 094106.

\bibitem{1817Davy}
H.~Davy, {Forms of miners' safety lamp}, Philos. T. R. Soc. A 107 (1817)
  77--86.

\bibitem{15NaKoLaBr}
P.~Nau, J.~Koppmann, A.~Lackner, A.~Brockhinke, {Detection of formaldehyde in
  flames using UV and MIR absorption spectroscopy}, Z. Phys. Chem. 229 (2015)
  483--494.

\bibitem{17FjHeBaLe}
P.~Fjodorow, O.~Hellmig, V.~M. Baev, H.~B. Levinsky, A.~V. Mokhov, {Intracavity
  absorption spectroscopy of formaldehyde from 6230 to 6420 cm$^{-1}$}, Appl.
  Phys. B 123 (2017) 147.

\bibitem{20DiPeStHa}
Y.~Ding, W.~Y. Peng, C.~L. Strand, R.~K. Hanson, {Quantitative measurements of
  broad-band mid-infrared absorption spectra of formaldehyde, acetaldehyde, and
  acetone at combustion-relevant temperatures near 5.7 $\mu$m}, J. Quant.
  Spectrosc. Rad. Transf. 248 (2020) 106981.

\bibitem{93KoAcKrMo}
O.~I. Korablev, M.~Ackerman, V.~A. Krasnopolsky, V.~I. Moroz, C.~Muller, A.~V.
  Rodin, S.~K. Aterya, {Tentative identification of formaldehyde in the Martian
  atmosphere}, Planet. Space Sci. 41 (1993) 441--451.

\bibitem{92BoCr}
D.~Bockelee-Morvan, J.~Crovisier, {Formaldehyde in comets. II. Excitation of
  the rotational lines}, Astron. Astrophys. 264 (1992) 282--291.

\bibitem{06MiReWoAb}
S.~N. Milam, A.~J. Remijan, M.~Womack, L.~Abrell, L.~M. Ziurys, S.~Wyckoff,
  A.~J. Apponi, D.~N. Friedel, L.~E. Snyder, J.~M. Veal, P.~Palmer, L.~M.
  Woodney, M.~F. A'Hearn, J.~R. Forster, M.~C.~H. Wright, I.~de~Pater, S.~Choi,
  M.~Gesmundo, {Formaldehyde in comets C/1995 O1 (Hale-Bopp), C/2002 T7
  (LINEAR), and C/2001 Q4 (NEAT): investigating the cometary origin of
  H$_2$CO}, Astrophys. J. 649 (2006) 1169.

\bibitem{11DeVeLiWe}
N.~Dello~Russo, R.~J. Vervack, Jr., C.~M. Lisse, H.~A. Weaver, H.~Kawakita,
  H.~Kobayashi, A.~L. Cochran, W.~M. Harris, A.~J. McKay, N.~Biver,
  D.~Bockel\'ee-Morvan, J.~Crovisier, {The volatile composition and activity of
  comet 103P/Hartley 2 during the EPOXI closest approach}, Astrophys. J. 734
  (2011) L8.

\bibitem{11ViMuDiBo}
G.~L. Villanueva, M.~J. Mumma, M.~A. Disanti, B.~P. Bonev, E.~L. Gibb,
  K.~Magee-Sauer, G.~A. Blake, C.~Salyk, {The molecular composition of Comet
  C/2007 W1 (Boattini): Evidence of a peculiar outgassing and a rich
  chemistry}, Icarus 216 (2011) 227--240.

\bibitem{14SaFoWaAl}
B.~A. Sargent, W.~Forrest, D.~M. Watson, P.~d'Alessio, N.~Calvet, E.~Furlan,
  K.~H. Kim, J.~Green, K.~Pontoppidan, I.~Richter, C.~Tayrien, {Emission from
  water vapor and absorption from other gases at 5--7.5 $\mu$m in Spitzer-IRS
  spectra of protoplanetary disks}, Astrophys. J. 792 (2014) 83.

\bibitem{70ZuBuPaSn}
B.~Zuckerman, D.~Buhl, P.~Palmer, L.~E. Snyder, {Observations of interstellar
  formaldehyde}, Astrophys. J. 160 (1970) 485.

\bibitem{93MaWoPl}
J.~G. Mangum, A.~Wootten, R.~L. Plambeck, {The physical structure of Orion-KL
  on 2500-au scales using the $K$-doublet transitions of formaldehyde},
  Astrophys. J. 409 (1993) 282--298.

\bibitem{80FoGoWiDo}
J.~R. Forster, W.~M. Goss, T.~L. Wilson, D.~Downes, H.~R. Dickel, {A
  formaldehyde maser in NGC7538}, Astron. Astrophys. 84 (1980) L1--L3.

\bibitem{07HoGoPa}
I.~M. Hoffman, W.~M. Goss, P.~Palmer, {The formaldehyde masers in Sgr B2: very
  long baseline array and very large array observations}, Astrophys. J. 654
  (2007) 971.

\bibitem{13WaZhGa}
J.-Z. Wang, Z.-Y. Zhang, Y.~Gao, {High resolution observations of the 6 cm
  H$_2$CO maser in NGC 6240}, Res. Astron. Astrophys. 13 (2013) 270--276.

\bibitem{14PaSo}
S.~Y. Parfenov, A.~M. Sobolev, {On the Class II methanol maser periodic
  variability due to the rotating spiral shocks in the gaps of discs around
  young binary stars}, Mon. Not. R. Astron. Soc. 444 (2014) 620--628.

\bibitem{21T2HoCl.H2CO}
J.~Terwisscha~van Scheltinga, M.~R. Hogerheijde, L.~I. Cleeves, R.~A. Loomis,
  C.~Walsh, K.~I. Oberg, E.~A. Bergin, J.~B. Bergner, G.~A. Blake, J.~K.
  Calahan, P.~Cazzoletti, E.~F. van Dishoeck, V.~V. Guzman, J.~Huang, M.~Kama,
  C.~Qi, R.~Teague, D.~J. Wilner, {The TW Hya Rosetta Stone Project. II.
  Spatially Resolved Emission of Formaldehyde Hints at Low-temperature
  Gas-phase Formation}, Astrophys. J. {906} ({2021}) 111.
\newblock \href {http://dx.doi.org/{10.3847/1538-4357/abc9ba}}
  {\path{doi:{10.3847/1538-4357/abc9ba}}}.

\bibitem{34DiKi}
G.~H. Dieke, G.~B. Kistiakowsky, {The structure of the ultraviolet absorption
  spectrum of formaldehyde. I}, Phys. Rev. 45 (1934) 4--28.

\bibitem{87ChFoMo}
M.-C. Chuang, M.~F. Foltz, C.~B. Moore, {$T_1$ barrier height, $S_1$--$T_1$
  intersystem crossing rate, and $S_0$ radical dissociation threshold for
  H$_2$CO, D$_2$CO, and HDCO}, J. Chem. Phys. 87 (1987) 3855.

\bibitem{51LaSt}
R.~B. Lawrance, M.~W.~P. Strandberg, {Centrifugal distortion in asymmetric top
  molecules. I. Ordinary formaldehyde, H$_2^{~12}$CO}, Phys. Rev. 83 (1951)
  363--369.

\bibitem{59TaShSh}
H.~Takuma, T.~Shimizu, K.~Shimoda, {Magnetic hyperfine spectrum of H$_2$CO by a
  maser}, J. Phys. Soc. Jpn. 14 (1959) 1595--1599.

\bibitem{60Oka}
T.~Oka, {Microwave spectrum of formaldehyde. II. Molecular structure in the
  ground state}, J. Phys. Soc. Jpn. 15 (1960) 2274--2279.

\bibitem{60OkHiSh}
T.~Oka, H.~Hirakawa, K.~Shimoda, {Microwave spectrum of formaldehyde. I.
  $K$-type doubling spectra}, J. Phys. Soc. Jpn. 15 (1960) 2265--2273.

\bibitem{60ShTaSh}
K.~Shimoda, H.~Takuma, T.~Shimizu, {Beam-type masers for radiofrequency
  spectroscopy}, J. Phys. Soc. Jpn. 15 (1960) 2036--2041.

\bibitem{63Esterowi}
L.~Esterowitz, {Rotational transitions and centrifugal distortion in uhf
  spectrum of formaldehyde}, J. Chem. Phys. 39 (1963) 247.

\bibitem{63ShKoTa}
T.~Shigenari, S.~Kobayashi, H.~Takuma, {(6.3) rotational spectrum of H$_2$CO by
  a radiofrequency beam-type maser}, J. Phys. Soc. Jpn. 18 (1963) 312--313.

\bibitem{64OkTaMo}
T.~Oka, K.~Takagi, Y.~Morino, {Microwave spectrum of formaldehyde in
  vibrationally excited states}, J. Mol. Spectrosc. 14 (1964) 27--52.

\bibitem{68Takami}
M.~Takami, {$K$-type doubling lines of H$_2$CO and HCOOH in the HF region}, J.
  Phys. Soc. Jpn. 24 (1968) 372--376.

\bibitem{69NaKaKuMo}
T.~Nakagawa, H.~Kashiwagi, H.~Kurihara, Y.~Morino, {Vibration-rotation spectra
  of formaldehyde}, J. Mol. Spectrosc. 31 (1969) 436--450.

\bibitem{70KrGeShPo}
A.~F. Krupnov, L.~I. Gershtein, V.~G. Shustrov, V.~V. Polyakov, {Submillimeter
  microwave spectroscopy of formaldehyde}, Opt. Spectrosc. 28 (1970) 257.

\bibitem{70TuThTo}
K.~D. Tucker, P.~Thaddeus, G.~R. Tomasevich, {Precise laboratory measurement of
  4830-MHz formaldehyde rotational transition}, Astrophys. J. 161 (1970)
  L153--L154.

\bibitem{71NaMo}
T.~Nakagawa, Y.~Morino, {Coriolis interactions in $\nu_4$ and $\nu_6$ bands of
  formaldehyde}, J. Mol. Spectrosc. 38 (1971) 84--106.

\bibitem{71TuToTh}
K.~D. Tucker, G.~R. Tomasevich, P.~Thaddeus, {Laboratory measurement of the 6
  centimeter formaldehyde transitions}, Astrophys. J. 169 (1971) 429--440.

\bibitem{72JoLoKi}
D.~R. Johnson, F.~J. Lovas, W.~H. Kirchhoff, {Microwave spectra of molecules of
  astrophysical interest: 1. Formaldehyde, formamide, and thioformaldehyde}, J.
  Phys. Chem. Ref. Data 1 (1972) 1011--1046.

\bibitem{72Nerf}
R.~B. Nerf, {Laboratory measurement of millimeter-wavelength spectrum of
  formaldehyde}, Astrophys. J. 174 (1972) 467--468.

\bibitem{72TuToTh}
K.~D. Tucker, G.~R. Tomasevich, P.~Thaddeus, {Laboratory measurement of the
  2-centimeter, $2_{11}-2_{12}$, transition of normal formaldehyde and its
  carbon-13 and oxygen-18 species}, Astrophys. J. 174 (1972) 463--466.

\bibitem{73ChGu}
J.~C. Chardon, D.~Guichon, {Structure hyperfine du spectre basse fréquence de
  H$_2$CO}, J. Phys. (Paris) 34 (1973) 791--802.

\bibitem{73ChFrJoOk}
F.~Y. Chu, S.~M. Freund, J.~W.~C. Johns, T.~Oka, {$\Delta K = 2$ transitions in
  H$_2$CO and D$_2$CO}, J. Mol. Spectrosc. 48 (1973) 328--335.

\bibitem{73JoMc}
J.~W.~C. Johns, A.~R.~W. McKellar, {Stark spectroscopy with the CO laser: The
  $\nu_2$ fundamentals of H$_2$CO and D$_2$CO}, J. Mol. Spectrosc. 48 (1973)
  354--371.

\bibitem{73Toth}
R.~A. Toth, {High resolution measurements of the line positions and strengths
  of the $2\nu_2$ band of H$_2$CO}, J. Mol. Spectrosc. 46 (1973) 470--489.

\bibitem{75Nerf}
R.~B. Nerf, {Pressure broadening and shift in the millimeter-wave spectrum of
  formaldehyde}, J. Mol. Spectrosc. 58 (1975) 451--473.

\bibitem{75TaYaNaKu}
K.~Tanaka, K.~Yamada, T.~Nakagawa, K.~Kuchitsu, J.~Overend, {Infrared spectrum
  of formaldehyde: Coriolis interactions in the combination bands $\nu_2$+
  $\nu_6$ and $\nu_2$+ $\nu_3$}, J. Mol. Spectrosc. 54 (1975) 243--260.

\bibitem{77AlJoMc}
M.~Allegrini, J.~Johns, A.~McKellar, {A study of the Coriolis-coupled $\nu_4$,
  $\nu_6$, and $\nu_3$ fundamental bands and the $\nu_5$ $\leftarrow$ $\nu_6$
  difference band of H$_2$CO; measurement of the dipole moment for $\nu_5 =
  1$}, J. Mol. Spectrosc. 67 (1977) 476--495.

\bibitem{77AlJoMcb}
M.~Allegrini, J.~W.~C. Johns, A.~R.~W. McKellar, {Stark spectroscopy with the
  CO laser: the $\nu_3$ fundamental band of H$_2$CO}, J. Mol. Spectrosc. 66
  (1977) 69--78.

\bibitem{77ChGu}
J.~C. Chardon, D.~Guichon, {Spectre radiofr\'equence de H$_2$CO dans des
  \'etats vibrationnels excit\'es}, J. Phys. (Paris) 38~(2) (1977) 113--120.

\bibitem{77FaKrMu}
B.~Fabricant, D.~Krieger, J.~S. Muenter, {Molecular beam electric resonance
  study of formaldehyde, thioformaldehyde, and ketene}, J. Chem. Phys. 67
  (1977) 1576--1586.

\bibitem{78DaWiBe}
D.~Dangoisse, E.~Willemot, J.~Bellet, {Microwave spectrum of formaldehyde and
  its isotopic species in D, $^{13}$C, and $^{18}$O: study of Coriolis
  resonance between $\nu_4$ and $\nu_6$ vibrational excited states}, J. Mol.
  Spectrosc. 71 (1978) 414--429.

\bibitem{78Pine}
A.~S. Pine, {Doppler-limited spectra of C--H stretching fundamentals of
  formaldehyde}, J. Mol. Spectrosc. 70 (1978) 167--178.

\bibitem{79BrHuPi}
L.~R. Brown, R.~H. Hunt, A.~S. Pine, {Wavenumbers, line strengths, and
  assignments in the {D}oppler-limited spectrum of formaldehyde from 2700
  cm$^{-1}$ to 3000 cm$^{-1}$}, J. Mol. Spectrosc. 75 (1979) 406--428.

\bibitem{79HaTi}
J.~L. Hardvick, S.~M. Till, {Laser excited resonance fluorescence in
  formaldehyde}, J. Chem. Phys. 70 (1979) 2340.

\bibitem{80CoWi}
R.~Cornet, G.~Winnewisser, {A precise study of the rotational spectrum of
  formaldehyde H$_2^{~12}$C$^{16}$O, H$_2^{~13}$C$^{16}$O,
  H$_2^{~12}$C$^{18}$O, H$_2^{~13}$C$^{18}$O}, J. Mol. Spectrosc. 80 (1980)
  438--452.

\bibitem{81ChMi}
J.-C. Chardon, J.-J. Miller, {Spectroscopie r\'esonance \'electrique des jets
  mol\'eculaires : formes des raies de résonance en présence d'effet
  Doppler}, Can. J. Phys. 59 (1981) 378--386.

\bibitem{81SwSa}
D.~M. Sweger, R.~L. Sams, {Diode laser spectra of the $\nu_2$ band of
  H$_2^{~12}$CO and H$_2^{~13}$CO}, J. Mol. Spectrosc. 87 (1981) 18--28.

\bibitem{82BrJoMcWo}
C.~Br\'echignac, J.~W.~C. Johns, A.~R.~W. McKellar, M.~Wong, {The $\nu_2$
  fundamental band of H$_2$CO}, J. Mol. Spectrosc. 96 (1982) 353--361.

\bibitem{85TiChKuHu}
T.~Tipton, J.-I. Choe, S.~G. Kukolich, R.~Hubbard, {Fourier transform
  spectroscopy on the $3\nu_2$, $2\nu_2 + \nu_6$ and $\nu_3 + \nu_5$ bands of
  H$_2$CO}, J. Mol. Spectrosc. 114 (1985) 239--256.

\bibitem{87NaDaRe}
S.~Nadler, S.~J. Daunt, D.~C. Reuter, {Tunable diode-laser measurements of
  formaldehyde foreign-gas broadening parameters and line strengths in the
  9--11 $\mu$m region}, Appl. Optics 26 (1987) 1641--1646.

\bibitem{88ClVa}
D.~S. Cline, P.~L. Varghese, {High resolution spectral measurements in the
  $\nu_5$ band of formaldehyde using a tunable IR diode laser}, Appl. Optics 27
  (1988) 3219--3224.

\bibitem{89ReNaDaJo}
D.~C. Reuter, S.~Nadler, S.~J. Daunt, J.~W.~C. Johns, {Frequency and intensity
  analysis of the $\nu_3$ band, $\nu_4$ band and $\nu_6$ band of formaldehyde},
  J. Chem. Phys. 91 (1989) 646--654.

\bibitem{94ItNaTa}
F.~Ito, T.~Nakanaga, H.~Takeo, {FTIR spectra of the $2\nu_4$, $\nu_4 + \nu_6$
  and $2\nu_6$ bands of formaldehyde}, Spectrochim. Acta A 50 (1994)
  1397--1412.

\bibitem{96BoDePoLi}
R.~Bocquet, J.~Demaison, L.~Poteau, M.~Liedtke, S.~Belov, K.~M.~T. Yamada,
  G.~Winnewisser, C.~Gerke, J.~Gripp, T.~K\"ohler, {The ground state rotational
  spectrum of formaldehyde}, J. Mol. Spectrosc. 177 (1996) 154--159.

\bibitem{03ThCaRiMu}
P.~Theul\'e, A.~Callegari, T.~R. Rizzo, J.~S. Muenter, {Fluorescence detected
  microwave Stark effect measurements in excited vibrational states of
  H$_2$CO}, J. Chem. Phys. 119 (2003) 8910--8915.

\bibitem{88NaReDaJo}
S.~Nadler, D.~C. Reuter, S.~J. Daunt, J.~W.~C. Johns, {The $\nu_3$, $\nu_4$ and
  $\nu_6$ bands of formaldehyde: a spectral catalog from 900 to 1580
  cm$^{-1}$}, Nasa technical memorandum, NASA (1988).

\bibitem{96BoHaGrSt}
R.~J. Bouwens, J.~A. Hammerschmidt, M.~M. Grzeskowiak, T.~A. Stegink, P.~M.
  Yorba, W.~F. Polik, {Pure vibrational spectroscopy of $S_0$ formaldehyde by
  dispersed fluorescence}, J. Chem. Phys. 104 (1996) 460--479.

\bibitem{96LuCoFrCr}
D.~Luckhaus, M.~J. Coffey, M.~D. Fritz, F.~F. Crim, {Experimental and
  theoretical vibrational overtone spectra of $v_{\rm CH}=3,4,5$, and 6 in
  formaldehyde (H$_2$CO)}, J. Chem. Phys. 104 (1996) 3472--3478.

\bibitem{02BaCoHaPe}
H.~Barry, L.~Corner, G.~Hancock, R.~Peverall, G.~A.~D. Ritchie, {Cross sections
  in the $2\nu_5$ band of formaldehyde studied by cavity enhanced absorption
  spectroscopy near 1.76 $\mu$m}, Phys. Chem. Chem. Phys. 4 (2002) 445--450.

\bibitem{03BrMuLeWi}
S.~Br\"unken, H.~S.~P. M\"uller, F.~Lewen, G.~Winnewisser, {High accuracy
  measurements on the ground state rotational spectrum of formaldehyde
  (H$_2$CO) up to 2 THz}, Phys. Chem. Chem. Phys. 5 (2003) 1515--1518.

\bibitem{03PeKeFl}
A.~Perrin, F.~Keller, J.~M. Flaud, {New analysis of the $\nu_2$, $\nu_3$,
  $\nu_4$, and $\nu_6$ bands of formaldehyde, H$_2^{~12}$C$^{16}$O line
  positions and intensities in the 5--10 $\mu$m spectral region}, J. Mol.
  Spectrosc. 221 (2003) 192--198.

\bibitem{05StGaVeRu}
M.~Staak, E.~W. Gash, D.~S. Venables, A.~A. Ruth, {The rotationally-resolved
  absorption spectrum of formaldehyde from 6547 to 6804 cm$^{-1}$}, J. Mol.
  Spectrosc. 229 (2005) 115--121.

\bibitem{06FlLaSaSh}
J.~M. Flaud, W.~J. Lafferty, R.~L. Sams, S.~W. Sharpe, {High resolution
  spectroscopy of H$_2^{~12}$C$^{16}$O in the 1.9 to 2.56 $\mu$m spectral
  range}, Mol. Phys. 104 (2006) 1891--1903.

\bibitem{06PeVaDa}
A.~Perrin, A.~Valentin, L.~Daumont, {New analysis of the $2\nu_4$,
  $\nu_4+\nu_6$, $2\nu_6$, $\nu_3+\nu_4$, $\nu_3+\nu_6$, $\nu_1$, $\nu_5$,
  $\nu_2+\nu_4$, $2\nu_3$, $\nu_2+\nu_6$ and $\nu_2+\nu_3$ bands of
  formaldehyde H$_2^{~12}$C$^{16}$O: line positions and intensities in the
  3.5$\mu$m spectral region}, J. Mol. Struct. 780-781 (2006) 28--44.

\bibitem{06PeBrUtHa}
R.~Perez, J.~M. Brown, Y.~Utkin, J.~Han, R.~F. Curl, {Observation of hot bands
  in the infrared spectrum of H$_2$CO}, J. Mol. Spectrosc. 236 (2006) 151--157.

\bibitem{07TcPeLa}
F.~K. Tchana, A.~Perrin, N.~Lacome, {New analysis of the $\nu_2$ band of
  formaldehyde (H$_2^{~12}$C$^{16}$O): line positions for the $\nu_2$, $\nu_3$,
  $\nu_4$ and $\nu_6$ interacting bands}, J. Mol. Spectrosc. 245 (2007)
  141--144.

\bibitem{07ZhGaDeHu}
W.~Zhao, X.~Gao, L.~Deng, T.~Huang, T.~Wu, W.~Zhang, {Absorption spectroscopy
  of formaldehyde at 1.573 $\mu$m}, J. Quant. Spectrosc. Rad. Transf. 107
  (2007) 331--339.

\bibitem{07SaBaHaRi}
S.~Saha, H.~Barry, G.~Hancock, G.~A.~D. Ritchie, C.~M. Western, {Rotational
  analysis of the $2\nu_5$ band of formaldehyde}, Mol. Phys. 105 (2007)
  797--805.

\bibitem{09MaPeJaBa}
L.~Margul\'es, A.~Perrin, R.~Janeckova, S.~Bailleux, C.~P. Endres, T.~F.
  Giesen, S.~Schlemmer, {Rotational transitions within the $2^1$, $3^1$, $4^1$,
  and $6^1$ states of formaldehyde H$_2^{~12}$C$^{16}$O}, Can. J. Phys. 87
  (2009) 425--435.

\bibitem{09PeJaTcLa}
A.~Perrin, D.~Jacquemart, F.~K. Tchana, N.~Lacome, {Absolute line intensities
  measurements and calculations for the 5.7 and 3.6 $\mu$m bands of
  formaldehyde}, J. Quant. Spectrosc. Rad. Transf. 110 (2009) 700--716.

\bibitem{09CiMaCi}
J.~Cihelka, I.~Matulkov\'a, S.~Civi\v{s}, {Laser diode photoacoustic and FTIR
  laser spectroscopy of formaldehyde in the 2.3 $\mu$m and 3.5 $\mu$m spectral
  range}, J. Mol. Spectrosc. 256 (2009) 68--74.

\bibitem{10JaLaTcGa}
D.~Jacquemart, A.~Laraia, F.~K. Tchana, R.~R. Gamache, A.~Perrin, N.~Lacome,
  {Formaldehyde around 3.5 and 5.7 $\mu$m: measurement and calculation of
  broadening coefficients}, J. Quant. Spectrosc. Rad. Transf. 111 (2010)
  1209--1222.

\bibitem{12ElCuGuHi}
S.~Eliet, A.~Cuisset, M.~Guinet, F.~Hindle, G.~Mouret, R.~Bocquet, J.~Demaison,
  {Rotational spectrum of formaldehyde reinvestigated using a photomixing THz
  synthesizer}, J. Mol. Spectrosc. 279 (2012) 12--15.

\bibitem{15RuHeHeFi}
A.~A. Ruth, U.~Heitmann, E.~Heinecke, C.~Fittschen, {The rotationally-resolved
  absorption spectrum of formaldehyde from 6547 to 7051 cm$^{-1}$}, Z. Phys.
  Chem. 229 (2015) 1609--1624.

\bibitem{17MuLe}
H.~S.~P. M\"uller, F.~Lewen, {Submillimeter spectroscopy of H$_2$C$^{17}$O and
  a revisit of the rotational spectra of H$_2$C$^{18}$O and H$_2$C$^{16}$O}, J.
  Mol. Spectrosc. 331 (2017) 28--33.

\bibitem{17TaAdNg}
T.~L. Tan, R.~\'Adawiah, L.~L. Ng, {The $2\nu_2$ bands of H$_2^{~12}$CO and
  H$_2^{~13}$CO by high-resolution FTIR spectroscopy}, J. Mol. Spectrosc. 340
  (2017) 16--20.

\bibitem{07FuCsTe}
T.~Furtenbacher, A.~G. Cs\'asz\'ar, J.~Tennyson, {MARVEL: measured active
  rotational-vibrational energy levels}, J. Mol. Spectrosc. 245 (2007)
  115--125.

\bibitem{12FuCs}
T.~Furtenbacher, A.~G. Cs\'asz\'ar, {MARVEL: measured active
  rotational-vibrational energy levels. II. Algorithmic improvements}, J.
  Quant. Spectrosc. Rad. Transf. 113 (2012) 929--935.

\bibitem{19ToFuTeCs}
R.~T\'obi\'as, T.~Furtenbacher, J.~Tennyson, A.~G. Cs\'asz\'ar, {Accurate
  empirical rovibrational energies and transitions of H$_2^{~16}$O}, Phys.
  Chem. Chem. Phys. 21 (2019) 3473--3495.

\bibitem{09TeBeBrCa}
J.~Tennyson, P.~F. Bernath, L.~R. Brown, A.~Campargue, M.~R. Carleer, A.~G.
  Cs\'asz\'ar, R.~R. Gamache, J.~T. Hodges, A.~Jenouvrier, O.~V. Naumenko,
  O.~L. Polyansky, L.~S. Rothman, R.~A. Toth, A.~C. Vandaele, N.~F. Zobov,
  L.~Daumont, A.~Z. Fazliev, T.~Furtenbacher, I.~E. Gordon, S.~N. Mikhailenko,
  S.~V. Shirin, {IUPAC critical evaluation of the rotational-vibrational
  spectra of water vapor. Part I. Energy levels and transition wavenumbers for
  H$_2^{~17}$O and H$_2^{~18}$O}, J. Quant. Spectrosc. Rad. Transf. 110 (2009)
  573--596.

\bibitem{12TeYu}
J.~Tennyson, S.~N. Yurchenko, {ExoMol: molecular line lists for exoplanet and
  other atmospheres}, Mon. Not. R. Astron. Soc. 425 (2012) 21--33.

\bibitem{15AlYuYaTe}
A.~F. Al-Refaie, S.~N. Yurchenko, A.~Yachmenev, J.~Tennyson, {ExoMol line lists
  VIII: a variationally computed line list for hot formaldehyde}, Mon. Not. R.
  Astron. Soc. 448 (2015) 1704--1714.

\bibitem{18Birkbyb}
J.~L. Birkby, {Spectroscopic direct detection of exoplanets}, in: H.~Deeg,
  J.~Belmonte (Eds.), Handbook of exoplanets, Springer, New York, 2018, pp.
  1485--1508.

\bibitem{15HoKoSnBr}
H.~J. Hoeijmakers, R.~J. de~Kok, I.~A.~G. Snellen, M.~Brogi, J.~L. Birkby,
  H.~Schwarz, {A search for TiO in the optical high-resolution transmission
  spectrum of HD 209458b: hindrance due to inaccuracies in the line database},
  Astron. Astrophys. 575 (2015) A20.

\bibitem{20TeYuAlCl}
J.~Tennyson, S.~N. Yurchenko, A.~F. Al-Refaie, V.~H.~J. Clark, K.~L. Chubb,
  E.~K. Conway, A.~Dewan, M.~N. Gorman, C.~Hill, A.~E. Lynas-Gray, T.~Mellor,
  L.~K. McKemmish, A.~Owens, O.~L. Polyansky, M.~Semenov, W.~Somogyi,
  G.~Tinetti, A.~Upadhyay, I.~Waldmann, Y.~Wang, S.~Wright, O.~P. Yurchenko,
  {The 2020 release of the ExoMol database: molecular line lists for exoplanet
  and other hot atmospheres}, J. Quant. Spectrosc. Rad. Transf. 255 (2020)
  107228.

\bibitem{11CsFu}
A.~G. Cs\'asz\'ar, T.~Furtenbacher, {Spectroscopic networks}, J. Mol.
  Spectrosc. 266 (2011) 99--103.

\bibitem{16CsFuAr}
A.~G. Cs\'asz\'ar, T.~Furtenbacher, P.~\'Arend\'as, {Small molecules -- big
  data}, J. Phys. Chem. A 120 (2016) 8949--8969.

\bibitem{16ArFuCs}
P.~\'Arend\'as, T.~Furtenbacher, A.~G. Cs\'asz\'ar, {On spectra of spectra}, J.
  Math. Chem. 54 (2016) 806--822.

\bibitem{20FuToTePo}
T.~Furtenbacher, R.~T\'obi\'as, J.~Tennyson, O.~L. Polyansky, A.~G.
  Cs\'asz\'ar, {W2020: A database of validated rovibrational experimental
  transitions and empirical energy levels of H$_2^{~16}$O}, J. Phys. Chem. Ref.
  Data 49 (2020) 033101.

\bibitem{14FuArMeCs}
T.~Furtenbacher, P.~\'Arend\'as, G.~Mellau, A.~G. Cs\'asz\'ar, {Simple
  molecules as complex systems}, Sci. Rep. 4 (2014) 4654.

\bibitem{20ToFuSiCs}
R.~T\'obi\'as, T.~Furtenbacher, I.~Simk\'o, A.~G. Cs\'asz\'ar, M.~L. Diouf,
  F.~M.~J. Cozijn, J.~M.~A. Staa, E.~J. Salumbides, W.~Ubachs,
  {Spectroscopic-network-assisted precision spectroscopy and its application to
  water}, Nat. Commun. 11 (2020) 1708.

\bibitem{10TeBeBrCa}
J.~Tennyson, P.~F. Bernath, L.~R. Brown, A.~Campargue, A.~G. Cs\'asz\'ar,
  L.~Daumont, R.~R. Gamache, J.~T. Hodges, O.~V. Naumenko, O.~L. Polyansky,
  L.~S. Rothman, R.~A. Toth, A.~C. Vandaele, N.~F. Zobov, S.~Fally, A.~Z.
  Fazliev, T.~Furtenbacher, I.~E. Gordon, S.-M. Hu, S.~N. Mikhailenko, B.~A.
  Voronin, {IUPAC critical evaluation of the rotational-vibrational spectra of
  water vapor. Part II. Energy levels and transition wavenumbers for
  HD$^{16}$O, HD$^{17}$O, and HD$^{18}$O}, J. Quant. Spectrosc. Rad. Transf.
  110 (2010) 2160--2184.

\bibitem{13TeBeBrCa}
J.~Tennyson, P.~F. Bernath, L.~R. Brown, A.~Campargue, A.~G. Cs\'asz\'ar,
  L.~Daumont, R.~R. Gamache, J.~T. Hodges, O.~V. Naumenko, O.~L. Polyansky,
  L.~S. Rothman, A.~C. Vandaele, N.~F. Zobov, A.~R.~A. Derzi, C.~F\'abri, A.~Z.
  Fazliev, T.~Furtenbacher, I.~E. Gordon, L.~Lodi, I.~I. Mizus, {IUPAC critical
  evaluation of the rotational-vibrational spectra of water vapor. Part III.
  Energy levels and transition wavenumbers for H$_2^{~16}$O}, J. Quant.
  Spectrosc. Rad. Transf. 117 (2013) 29--80.

\bibitem{14TeBeBrCa}
J.~Tennyson, P.~F. Bernath, L.~R. Brown, A.~Campargue, A.~G. Cs\'asz\'ar,
  L.~Daumont, R.~R. Gamache, J.~T. Hodges, O.~V. Naumenko, O.~L. Polyansky,
  L.~S. Rothman, A.C.Vandaele, N.~F. Zobov, N.~D\'enes, A.~Z. Fazliev,
  T.~Furtenbacher, I.~E. Gordon, S.-M.Hu, T.~Szidarovszky, I.~A. Vasilenko,
  {IUPAC critical evaluation of the rotational-vibrational spectra of water
  vapor. Part IV. Energy levels and transition wavenumbers for D$_2^{~16}$O,
  D$_2^{~17}$O, and D$_2^{~18}$O}, J. Quant. Spectrosc. Rad. Transf. 117 (2014)
  93--108.

\bibitem{14TeBeBrCab}
J.~Tennyson, P.~F. Bernath, L.~R. Brown, A.~Campargue, A.~G. Cs\'asz\'ar,
  L.~Daumont, R.~R. Gamache, J.~T. Hodges, O.~V. Naumenko, O.~L. Polyansky,
  L.~S. Rothman, A.~C. Vandaele, N.~F. Zobov, {A database of water transitions
  from experiment and theory (IUPAC technical report)}, Pure Appl. Chem. 86
  (2014) 71--83.

\bibitem{16FuSzCsBe}
T.~Furtenbacher, I.~Szab\'o, A.~G. Cs\'asz\'ar, P.~F. Bernath, S.~N. Yurchenko,
  J.~Tennyson, {Experimental energy levels and partition function of the
  $^{12}$C$_2$ molecule}, Astrophys. J. Suppl. S. 224 (2016) 44.

\bibitem{17McMaShSa}
L.~K. McKemmish, T.~Masseron, S.~Sheppard, E.~Sandeman, Z.~Schofield,
  T.~Furtenbacher, A.~G. Cs\'asz\'ar, J.~Tennyson, C.~Sousa-Silva, {MARVEL
  analysis of the measured high-resolution rovibronic spectra of
  $^{48}$Ti$^{16}$O}, Astrophys. J. Suppl. S. 228 (2017) 15.

\bibitem{18McBoGoSh}
L.~K. McKemmish, J.~Borsovszky, K.~L. Goodhew, S.~Sheppard, A.~F.~V. Bennett,
  A.~D.~J. Martin, A.~Singh, C.~A.~J. Sturgeon, T.~Furtenbacher, A.~G.
  Cs\'asz\'ar, J.~Tennyson, {MARVEL analysis of the measured high-resolution
  rovibronic spectra of $^{90}$Zr$^{16}$O}, Astrophys. J. 867 (2018) 33.

\bibitem{19DaShJoKh}
D.~Darby-Lewis, H.~Shah, D.~Joshi, F.~Khan, M.~Kauwo, N.~Sethi, P.~F. Bernath,
  T.~Furtenbacher, R.~T\'obi\'as, A.~G. Cs\'asz\'ar, J.~Tennyson, {MARVEL
  analysis of the measured high-resolution spectra of $^{14}$NH}, J. Mol.
  Spectrosc. 362 (2019) 69--76.

\bibitem{20McSyBoYu}
L.~K. McKemmish, A.-M. Syme, J.~Borsovszky, S.~N. Yurchenko, J.~Tennyson,
  T.~Furtenbacher, A.~G. Cs\'asz\'ar, {An update to the MARVEL data set and
  ExoMol line list for $^{12}$C$_2$}, Mon. Not. R. Astron. Soc. 497 (2020)
  1081--1097.

\bibitem{13FuSzFaCs}
T.~Furtenbacher, T.~Szidarovszky, C.~F\'abri, A.~G. Cs\'asz\'ar, {MARVEL
  analysis of the rotational-vibrational states of the molecular Ions
  H$_2$D$^+$ and D$_2$H$^+$}, Phys. Chem. Chem. Phys. 15 (2013) 10181--10193.

\bibitem{13FuSzMaFa}
T.~Furtenbacher, T.~Szidarovszky, E.~M\'atyus, C.~F\'abri, A.~G. Cs\'asz\'ar,
  {Analysis of the rotational-vibrational states of the molecular ion H$_3^+$},
  J. Chem. Theor. Comput. 9 (2013) 5471--5478.

\bibitem{18ChNaKeBa}
K.~L. Chubb, O.~Naumenko, S.~Keely, S.~Bartolotto, S.~MacDonald, M.~Mukhtar,
  A.~Grachov, J.~White, E.~Coleman, A.~Liu, A.~Z. Fazliev, E.~R. Polovtseva,
  V.-M. Horneman, A.~Campargue, T.~Furtenbacher, A.~G. Cs\'asz\'ar, S.~N.
  Yurchenko, J.~Tennyson, {MARVEL analysis of the measured high-resolution
  rovibrational spectra of H$_2$S}, J. Quant. Spectrosc. Rad. Transf. 218
  (2018) 178--186.

\bibitem{18ToFuCsNa}
R.~T\'obi\'as, T.~Furtenbacher, A.~G. Cs\'asz\'ar, O.~V. Naumenko, J.~Tennyson,
  J.-M. Flaud, P.~Kumar, B.~Poirier, {Critical evaluation of measured
  rotational-vibrational transitions of four sulphur isotopologues of
  S$^{16}$O$_2$}, J. Quant. Spectrosc. Rad. Transf. 208 (2018) 152--163.

\bibitem{15AlFuYuTe}
A.~R. {Al Derzi}, T.~Furtenbacher, S.~N. Yurchenko, J.~Tennyson, A.~G.
  Cs\'asz\'ar, {MARVEL analysis of the measured high-resolution spectra of
  $^{14}$NH$_3$}, J. Quant. Spectrosc. Rad. Transf. 161 (2015) 117--130.

\bibitem{18ChJoFrCh}
K.~L. Chubb, M.~Joseph, J.~Franklin, N.~Choudhury, T.~Furtenbacher, A.~G.
  Cs\'asz\'ar, G.~Gaspard, P.~Oguoko, A.~Kelly, S.~N. Yurchenko, J.~Tennyson,
  C.~Sousa-Silva, {MARVEL analysis of the measured high-resolution spectra of
  C$_2$H$_2$}, J. Quant. Spectrosc. Rad. Transf. 204 (2018) 42--55.

\bibitem{20FuCoTeYu}
T.~Furtenbacher, P.~A. Coles, J.~Tennyson, S.~N. Yurchenko, S.~Yu, B.~Drouin,
  R.~T\'obi\'as, A.~G. Cs\'asz\'ar, {Empirical rovibrational energy levels of
  ammonia up to 7500 cm$^{-1}$}, J. Quant. Spectrosc. Rad. Transf. 251 (2020)
  107027.

\bibitem{11FaMaFuNe}
C.~F\'abri, E.~M\'atyus, T.~Furtenbacher, L.~Nemes, B.~Mih\'aly, T.~Zolt\'ani,
  A.~G. Cs\'asz\'ar, {Variational quantum mechanical and active database
  approaches to the rotational-vibrational spectroscopy of ketene, H$_2$CCO},
  J. Chem. Phys. 135 (2011) 094307.

\bibitem{55Mulliken}
R.~S. Mulliken, {Report on notation for the spectra of polyatomic molecules},
  J. Chem. Phys. 23 (1955) 1997--2011.

\bibitem{92Kroto}
H.~W. Kroto, {Molecular rotation spectra}, Dover, New York, 1992.

\bibitem{19MeCoEsDo}
M.~Melosso, B.~Conversazioni, C.~Degli~Esposti, L.~Dore, E.~Can\'e,
  F.~Tamassia, L.~Bizzocchi, {The pure rotational spectrum of $^{15}$ND$_2$
  observed by millimetre and submillimetre-wave spectroscopy}, J. Quant.
  Spectrosc. Rad. Transf. 222 (2019) 186--189.

\bibitem{20EsMeBiTa}
C.~Degli~Esposti, M.~Melosso, L.~Bizzocchi, F.~Tamassia, L.~Dore, Determination
  of a semi-experimental equilibrium structure of 1-phosphapropyne from
  millimeter-wave spectroscopy of {CH$_3$CP} and {CD$_3$CP}, J. Mol. Struct.
  1203 (2020) 127429.

\bibitem{64Lamb}
W.~E. {Lamb Jr.}, {Theory of an optical maser}, Phys. Rev. 134 (1964) A1429.

\bibitem{20MeDoGaPu}
M.~Melosso, L.~Dore, J.~Gauss, C.~Puzzarini, {Deuterium hyperfine splittings in
  the rotational spectrum of NH$_2$D as revealed by Lamb-dip spectroscopy}, J.
  Mol. Spectrosc. (2020) 111291.

\bibitem{97CaHaDe}
S.~Carter, N.~C. Handy, J.~Demaison, {The rotational levels of the ground
  vibrational state of formaldehyde}, Mol. Phys. 90 (1997) 729--737.

\bibitem{66TaEvSh}
H.~Takuma, K.~M. Evenson, T.~Shigenari, {Zeeman effect and magnetic hyperfine
  structure in low frequency transitions of H$_2$CO}, J. Phys. Soc. Jpn. 21
  (1966) 1622--1623.

\bibitem{56Erlandss}
G.~Erlandsson, {Millimeter wave spectrum of formaldehyde}, J. Chem. Phys. 25
  (1956) 579--580.

\bibitem{76NaYaKu}
T.~Nakagawa, K.~Yamada, K.~Kuchitsu, {Vibration-rotation spectrum of
  formaldehyde: C--H stretching fundamentals $\nu_1$ and $\nu_5$}, J. Mol.
  Spectrosc. 63 (1976) 485--508.

\bibitem{16GoPoBoEr}
I.~E. Gordon, M.~R. Potterbusch, D.~Bouquin, C.~C. Erdmann, J.~S. Wilzewski,
  L.~S. Rothman, {Are your spectroscopic data being used?}, J. Mol. Spectrosc.
  327 (2016) 232--238.

\bibitem{17GoRoHiKo}
I.~E. Gordon, L.~S. Rothman, C.~Hill, R.~V. Kochanov, Y.~Tan, P.~F. Bernath,
  M.~Birk, V.~Boudon, A.~Campargue, K.~V. Chance, B.~J. Drouin, J.-M. Flaud,
  R.~R. Gamache, J.~T. Hodges, D.~Jacquemart, V.~I. Perevalov, A.~Perrin, K.~P.
  Shine, M.-A.~H. Smith, J.~Tennyson, G.~C. Toon, H.~Tran, V.~G. Tyuterev,
  A.~Barbe, A.~G. Cs\'asz\'ar, V.~M. Devi, T.~Furtenbacher, J.~J. Harrison,
  J.-M. Hartmann, A.~Jolly, T.~J. Johnson, T.~Karman, I.~Kleiner, A.~A.
  Kyuberis, J.~Loos, O.~M. Lyulin, S.~T. Massie, S.~N. Mikhailenko,
  N.~Moazzen-Ahmadi, H.~S.~P. M{\"u}ller, O.~V. Naumenko, A.~V. Nikitin, O.~L.
  Polyansky, M.~Rey, M.~Rotger, S.~W. Sharpe, K.~Sung, E.~Starikova, S.~A.
  Tashkun, J.~V. Auwera, G.~Wagner, J.~Wilzewski, P.~Wcis\l{}o, S.~Yu, E.~J.
  Zak, {The HITRAN 2016 molecular spectroscopic database}, J. Quant. Spectrosc.
  Rad. Transf. 203 (2017) 3--69.

\bibitem{15National}
{National Academies of Sciences, Engineering, and Medicine}, {Frequency
  allocations and spectrum protection for scientific uses}, National Academies
  Press, Washington, 2015.

\bibitem{13DoHiYuTe}
M.~J. Down, C.~Hill, S.~N. Yurchenko, J.~Tennyson, L.~R. Brown, I.~Kleiner,
  {Re-analysis of ammonia spectra: Updating the HITRAN $^{14}$NH$_3$ database},
  J. Quant. Spectrosc. Rad. Transf. 130 (2013) 260--272.

\bibitem{20LaScTe}
V.~Laporta, I.~F. Schneider, J.~Tennyson, {Dissociative electron attachment
  cross sections for vibrationally excited NO molecule and N$^-$ anion
  formation}, Plasma Sources Sci. Technol. 20 (2020) 10LT01.

\bibitem{13TeHiYu}
J.~Tennyson, C.~Hill, S.~N. Yurchenko, {Data structures for ExoMol: Molecular
  line lists for exoplanet and other atmospheres}, in: {6$^{\rm th}$
  international conference on atomic and molecular data and their applications
  ICAMDATA-2012}, Vol. 1545 of AIP Conference Proceedings, AIP, New York, 2013,
  pp. 186--195.

\bibitem{18PoKyZoTe}
O.~L. Polyansky, A.~A. Kyuberis, N.~F. Zobov, J.~Tennyson, S.~N. Yurchenko,
  L.~Lodi, {ExoMol molecular line lists XXX: A complete high-accuracy line list
  for water}, Mon. Not. R. Astron. Soc. 480 (2018) 2597--2608.

\bibitem{18YuWiLeLo}
S.~N. Yurchenko, H.~Williams, P.~C. Leyland, L.~Lodi, J.~Tennyson, {ExoMol line
  lists XXVIII: the rovibronic spectrum of AlH}, Mon. Not. R. Astron. Soc. 479
  (2018) 1401--1411.

\bibitem{19McMaHoPe}
L.~K. McKemmish, T.~Masseron, H.~J. Hoeijmakers, V.~P\'erez-Mesa, S.~L. Grimm,
  S.~N. Yurchenko, J.~Tennyson, {ExoMol Molecular line lists -- XXXIII. The
  spectrum of titanium oxide}, Mon. Not. R. Astron. Soc. 488 (2019) 2836--2854.

\bibitem{20YuMeFrTe}
S.~N. Yurchenko, T.~M. Mellor, R.~S. Freedman, J.~Tennyson, {ExoMol molecular
  line lists XXXIX: ro-vibrational molecular line list for CO$_2$}, Mon. Not.
  R. Astron. Soc. 496 (2020) 5282--5291.

\bibitem{20YuTeMiMe}
S.~N. Yurchenko, J.~Tennyson, S.~Miller, V.~V. Melnikov, J.~O'Donoghue,
  L.~Moore, {ExoMol molecular line lists XL: ro-vibrational molecular line list
  for the hydronium ion (H$_3$O$^+$)}, Mon. Not. R. Astron. Soc. 497 (2020)
  2340--2351.

\bibitem{19CoYuTe}
P.~A. Coles, S.~N. Yurchenko, J.~Tennyson, {ExoMol molecular line lists XXXV: a
  rotation-vibration line list for hot ammonia}, Mon. Not. R. Astron. Soc. 490
  (2019) 4638--4647.

\end{thebibliography}

\end{document}